\documentclass[fleqn,usenatbib]{mnras}

\usepackage[T1]{fontenc}
\usepackage{ae,aecompl}

\usepackage{eso-pic}

\usepackage{graphicx}	
\usepackage{amsmath}	
\usepackage{amssymb}	
\usepackage[shortlabels]{enumitem}

\usepackage{multicol}

\title[IMF and systematics]{Stellar population analysis of MaNGA early-type galaxies:  IMF dependence and systematic effects}
\author[Bernardi et al.]{
\parbox{\textwidth}{
 M.~Bernardi$^{1}$\thanks{E-mail: \texttt{\rm \texttt{bernardm@sas.upenn.edu}}}, 
 H.~Dom\'{i}nguez S{\'a}nchez$^{2}$,
 R.~K.~Sheth$^{1}$,
 J.~R.~Brownstein$^{3}$ and
 R. R. Lane$^{4}$} \\
 \vspace{0.cm}\\~\\
$^{1}$ Department of Physics and Astronomy, University of Pennsylvania, Philadelphia, PA 19104, USA\\
 $^{2}$ Institute of Space Sciences (ICE, CSIC), Campus UAB, Carrer de Magrans, E-08193 Barcelona, Spain\\
 $^{3}$ Department of Physics and Astronomy, University of Utah, 115 S. 1400 E., Salt Lake City, UT 84112, USA\\
 $^{4}$ Centro de Investigaci{\'o}n en Astronom{\'i}a, Universidad Bernardo O'Higgins, Avenida Viel 1497, Santiago, Chile\\
}

\pubyear{2022}

\begin{document}
\label{firstpage}
\pagerange{\pageref{firstpage}--\pageref{lastpage}}
\maketitle

\begin{abstract}
We study systematics associated with estimating simple stellar population (SSP) parameters -- age, metallicity [M/H], $\alpha$-enhancement [$\alpha$/Fe] and IMF shape -- and associated $M_*/L$ gradients, of elliptical slow rotators (E-SRs), fast rotators (E-FRs) and S0s from stacked spectra of galaxies in the MaNGA survey.  These systematics arise from
   (i) how one normalizes the spectra when stacking; 
   (ii) having to subtract emission before estimating absorption line strengths; 
   (iii) the decision to fit the whole spectrum or just a few absorption lines;
   (iv) SSP model differences (e.g. isochrones, enrichment, IMF). 
The MILES+Padova SSP models, fit to the H$_\beta$, $\langle$Fe$\rangle$, TiO$_{\rm 2SDSS}$ and [MgFe] Lick indices in the stacks, indicate that out to the half-light radius $R_e$:
    (a) ages are younger and [$\alpha$/Fe] values are lower in the central regions but the opposite is true of [M/H];
    (b) the IMF is more bottom-heavy in the center, but is close to Kroupa beyond about $R_e/2$;
    (c) this makes $M_*/L$ about $2\times$ larger in the central regions than beyond $R_e/2$.
While the models of Conroy et al. (2018) return similar [M/H] and [$\alpha$/Fe] profiles, the age and (hence) $M_*/L$ profiles can differ significantly even for solar abundances and a Kroupa IMF; different responses to non-solar abundances and IMF parametrization further compound these differences. There are clear (model independent) differences between E-SRs, E-FRs and S0s: younger ages and less enhanced [$\alpha$/Fe] values suggest that E-FRs and S0s are not SSPs, but relaxing this assumption is unlikely to change their inferred $M_*/L$ gradients significantly.  
\end{abstract}




\section{Introduction}\label{sec:intro}

We are interested in quantifying gradients in the stellar populations of early-type galaxies.  These inform models of how such galaxies formed and assembled their stellar mass \cite[e.g., references in][]{C16,Smith2020}.  In addition, gradients affect the estimation and interpretation of observed correlations between the structural parameters of a galaxy and its mass.  As \cite{Bernardi2018b} have emphasized, gradients in the stellar initial mass function (IMF) must be quantified to accurately compare the stellar and dynamical masses of galaxies.
As we discuss in some detail in a companion paper \citep{Bernardi2022}, IMF gradients also modify the constraints one derives from the observed evolution of the half-light radii \cite[e.g.][]{Daddi2005, Mowla2019, Suess2019} of these galaxies.  The main goal of this paper is to enable such studies by determining stellar population, and especially IMF gradients, in a large sample of present day galaxies.

\begin{figure*}
    \centering
    \includegraphics[width=0.95\linewidth]{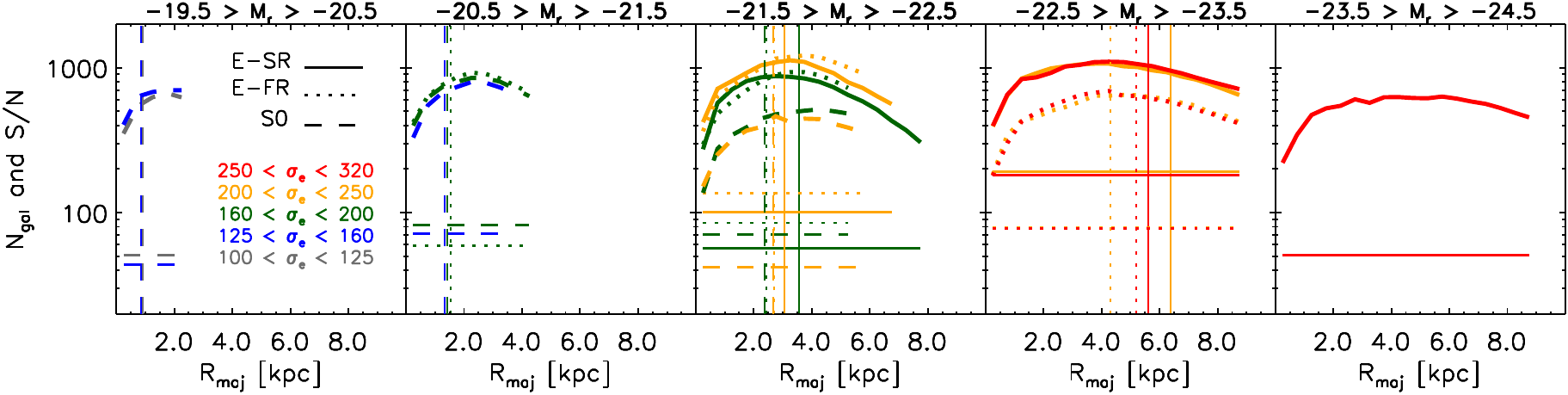}
    \caption{Number of galaxies which contribute to each stacked spectrum (horizontal lines in each panel), and resulting signal-to-noise ratio of the stack, which depends on the number of spaxels and the S/N of each spaxel (curves).  Panels show different luminosity bins, colors show different velocity dispersion bins, line styles show the morphological type (solid, dotted and dashed lines show slow rotator ellipticals E-SRs, fast rotator ellipticals E-FRs and S0s, respectively), and vertical lines show the median $R_e/2$ (where $R_e$ is the half-light radius in the $r$-band) of the galaxies which contribute to each stack.  Only stacks with $N_{\rm gal}\ge 40$ are used in our analysis to avoid cases in which a few objects dominate the stack. This results in composite spectra with very high S/N.}
    \label{fig:SN}
\end{figure*}

To quantify gradients we must first estimate stellar population parameters from the measured spectra.  This requires high signal to noise spectra, and models with which to interpret the measurements.  Unfortunately, such spectra are not available for large samples of individual objects.  Therefore, as we describe in previous papers of this series \citep[e.g.][]{DS2019}, we achieve high S/N by stacking spectra of objects having similar luminosities and central velocity dispersions.  One goal of this paper is to address some of the systematics associated with constructing such stacks.  

A second goal is to explore how systematic differences in the models used to interpret the stacked spectra propagate to the estimated stellar population gradients.  As our fiducial model we use the MILES models of \cite{Vazdekis2015} with the Padova isochrones.  These allow one to include some non-solar element abundances self-consistently, for a wide range of IMF shapes.  In \cite{DS2019} we motivated this choice by comparing with the models of \cite{TMJ2011} and \cite{TW2017}.  Here we compare our fiducial results with others that are based on the models of \cite{Conroy2018} (hereafter C18).  In principle, the C18 models are particularly interesting because they allow one to enhance elements one at a time, thereby enabling greater flexibility in modeling the stellar population.  In practice, the MILES and C18 models predict different spectra even when the abundance patterns are fixed to solar, so it is possible that SSP gradients determined using MILES will differ from those returned by C18.   We would like to quantify these differences.  

To do so, we assume that galaxies are effectively single-burst Simple Stellar Populations (SSPs) characterised by a few parameters:  age, metallicity, non-solar element abundances, and an IMF.  In principle, these can be estimated by fitting to the full spectrum \citep{Conroy2018, Vaughan18, FeldmeierKrause2021, Lonoce2021}.  In practice, this can be compromised by broadband variations which may be caused by dust, flux calibration issues, and so on.  A simple diagnostic of the latter is if the colors measured from the photometry disagree with those from the spectra.  In Section~\ref{sec:data} we show that this is indeed the case in the MaNGA dataset we use.  
Therefore, we consider how to proceed if we use a few Lick absorption lines, which were designed to be much less susceptible to broad band variations \citep{Worthey1994, Trager1998}.  
Our fiducial set includes the H$_\beta$, $\langle$Fe$\rangle$, [MgFe] and TiO$_{\rm 2SDSS}$ Lick absorption line indices \cite[e.g.][]{GonzalezPhD, LaBarbera2013, DS2019}. We discuss why this set of line indices is our preferred choice and some systematics associated with using other lines in Appendix~A.
Systematics associated with how one normalizes spectra when stacking, or from the correction for emission that is necessary for some absorption lines, are also discussed in Section~\ref{sec:data}.

Systematic differences between the MILES and C18 models, including how the IMF is parameterized, are considered in Section~\ref{sec:models}, where we motivate our fiducial choices of lines and models.  SSP gradients associated with our fiducial choices are presented and discussed in Section~\ref{sec:miles}, where we also compare the broadband colors predicted by our inferred SSP parameters with the observed colors (from imaging and spectra) to provide a consistency check.  A final section summarizes our results.  Two Appendices provide technical details associated with how various absorption lines and broadband colors respond to non-solar abundance patterns and IMFs.  

When estimating SSP $M_*/L$ values, we follow common practice \citep{Vazdekis2015, Conroy2018} by setting $M_*$ equal to the mass in stars that are still on the main sequence plus the mass in remnants (white dwarfs, neutron stars, black holes); we do {\em not} include the mass in (enriched) gas.  In addition, most of the $M_*/L$ values we quote are for the SDSS-$r$ band.  I.e., we work with $M_*/L_r$ in units of $M_\odot/L_{\odot,r}$.

\section{Data}\label{sec:data}

\subsection{MaNGA survey:  photometry and morphology}
The MaNGA survey (\citealt{Bundy2015,Drory2015,Law2015,Law2016,Yan2016a, Yan2016b, Aguado2019, Westfall2019}), which is a component of the Sloan Digital Sky Survey IV (\citealt{Gunn2006, Smee2013, Blanton2017}; hereafter SDSS IV), uses integral field units (IFUs) to measure spectra of nearby galaxies. The MaNGA final data release (DR17 -- \citealt{sdssDR17}) includes $\sim$ 10000 nearby ($0.03<z<0.15$) galaxies. The MaNGA selection function, while complicated, is well defined \citep{Wake2017}.  In what follows, when we refer to `weighted' quantities, the weight is {\tt ESWEIGHT} from \cite{Wake2017}, and is supposed to account for how the sample was selected.

In this work, the photometric and morphological information we use comes from two Value Added Catalogues that are included in the final MaNGA DR17 data release. The PyMorph Photometric Value Added Catalogue (DR17-MPP-VAC) provides photometric parameters from S\'ersic and S\'ersic + Exponential fits (we use the `truncated' magnitudes and sizes), and the Morphology Deep Learning Catalog (DR17-MMDL-VAC) provides morphological classifications.
We refer the reader to the corresponding references \citep{Fischer2019,DS21} for further details. 
These catalogs are updated/completed versions of the corresponding MaNGA DR15 VACs \citep{Fischer2019} and include 10,293 entries which correspond to 10,127 unique galaxies.
As described in Section 2.1 of \cite{Bernardi2022}, this yields 730 and 698 slow- and fast-rotator ellipticals (E-SRs and E-SRs), and 751 S0s.  
\cite{DS2019} recommend restricting the sample to $z\le 0.08$, so as to reduce aperture and evolution effects. Nevertheless, none of the results which follow are affected by the inclusion of objects with $0.08\le z<0.15$, so we simply work with the larger sample.

\subsection{Stellar population parameters}
It is generally accepted that fitting SSPs to the spectra of individual early type galaxies is reasonable.  However, this requires higher S/N spectra than we have available.  To increase S/N, we stack spectra of similar objects and then fit simple stellar population models (SSPs) to features in these stacked spectra.  Of course, even if an SSP is a good approximation for a single galaxy, it may not be appropriate for a stacked spectrum if there is substantial scatter in stellar population parameters of the objects which contribute to the stack.  (Likewise, if there are strong stellar population gradients in a galaxy, then treating it as an SSP -- as is traditionally done -- is also unreasonable.)  Therefore, we stack spectra of similar objects, like we did in our DR15 papers \citep{DS2019, Bernardi2019, DS2020}.  

Briefly, for each morphological type, we separate galaxies into luminosity bins which run from $M_r=-19.5$ to $-24.5$~mags in steps of 1~mag.  In each bin we further subdivide the objects based on the velocity dispersion measured within 0.1 of the half-light radius in bins which run from $\sigma_0 = 100$ to $\sigma_0 = 320$~kms$^{-1}$ (using the values reported in the MaNGA database) in steps of 0.1~dex \cite[see Figure~1 in][]{Bernardi2019}.  We then define a series of concentric elliptical annuli that are aligned with the main axis of the image and having axis ratios $b/a$, and stack together all the spaxels in each annulus.  The width of each annulus along the main axis is 0.5~kpc; finer bins do not have sufficient S/N, and the typical seeing radius is about 1~kpc anyway.  Figure~\ref{fig:SN} shows the number of galaxies and the resulting signal-to-noise ratio of each stack \cite[also see Table~1 in][]{Bernardi2022}.  

While this is very similar to our previous work \citep{DS2019} there are a few differences.
\begin{itemize}
\item The sample is twice as large.
\item Data Analysis Pipeline (DAP): We now use DR17 DAP HYB10-MILESHC-MASTARHC2 -- as we show below (Section~\ref{sec:emission}), the DR17 DAP HYB10-MILESHC-MASTARSSP tends to over-estimate the H$_\beta$ emission line while the DR15 DAP HYB10-GAU-MILESHC tended to under-estimate it.  As we discuss shortly, this is an important change.
\item We work with de-reddened spectra (i.e. corrected for Galactic extinction).  This does not affect Lick indices, but it does matter for estimating colors from the SED.
\item Discreteness effects:  We only use bins which include at least 40 galaxies to avoid cases in which a few objects dominate the stack.
\item We do not remove objects with $z > 0.08$.  There are few such objects at low luminosities, but approximately half of the most luminous objects lie at $z>0.08$.    Results are robust to their inclusion.
\item Use of unscaled radii:  As the scatter in $R_e$ at fixed $L$ and $\sigma_0$ is quite small, for each $L$ and $\sigma_0$ bin, we stack spectra in radial bins which are defined in kpc (i.e. we do not scale $R$ by $R_e$ prior to stacking in elliptical bins). As a result, uncertainties in the $R_e$ determination do not propagate into our stacks.  Results are robust to stacking in $R/R_e$ instead.
\end{itemize}  

\begin{figure}
    \centering
    \includegraphics[width=0.9\linewidth]{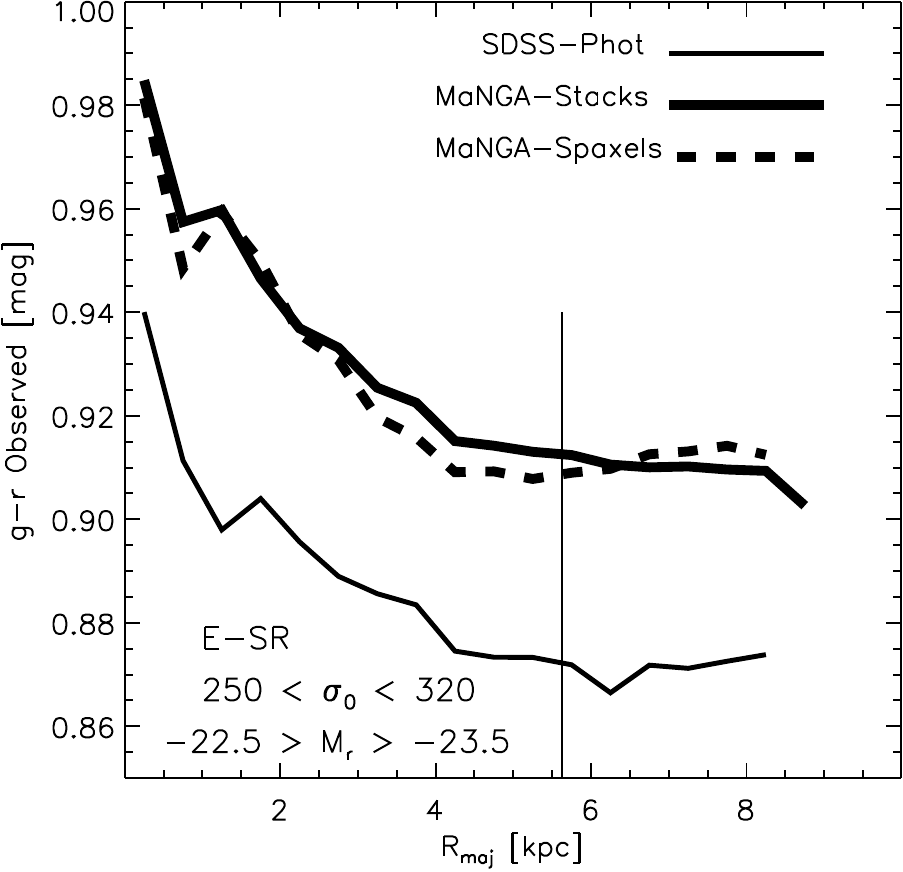}
    \caption{Comparison of $g-r$ color profiles measured from photometry (thin solid) and two estimates from the spectrum (thick solid or thick dashed) for E-SRs having the $L$ and $\sigma_0$ values shown.  The offset between the thick and thin curves complicates methods which use the full shape of the spectrum to constrain SSPs.  Vertical line shows the median $R_e/2$.}
    \label{fig:gmr1by1}
\end{figure}

\subsection{Colors from photometry vs spectroscopy}\label{sec:SED}
We begin with a comparison of the colors expected from the spectra with those in the photometry.  To simplify the comparison, we have chosen the E-SRs with $-23.5 > M_r > -24.5$ and $250<\sigma_0<320$~km/s.  We do not show them here, but we find qualitatively similar results in the other $L$ and $\sigma_0$ bins (also see Section~\ref{sec:colors}).

We have estimated colors from the spectra in two different ways.  In the first, for each spaxel of each object, we estimate the $g$- and $r$- band light by summing over wavelength, weighting by the SDSS filter curves.  We then plot the $g-r$ color versus scale $R_{\rm maj}$ (i.e. distance from galaxy center), such that all spaxels in the same ellipse of axis ratio $b/a$ and semimajor axis $R_{\rm maj}$ appear at $R_{\rm maj}$ (as opposed to the `circularized' radius $R_{\rm maj}\sqrt{b/a}$).  The thick dashed curve (MaNGA-Spaxels) in Figure~\ref{fig:gmr1by1} shows these measurements.  We compare this `observed color profile' with that from the photometry, before we discuss the thick solid curve.  

For the photometry, we do not work with the images themselves, but with the PyMorph fits in the $g$- and $r$- bands which have been corrected for seeing.  Since the spectra have not been corrected for seeing, we convolve the PyMorph fits with a Gaussian kernel that mimics the effects of seeing in the $g$-band (FWHM of about 1.4~arcsec), and estimate the light in the $g$- and $r$-bands that lies in a series of concentric elliptical annuli (axis ratio $b/a$).  (We have checked that this agrees with the $g-r$ color obtained directly from the images themselves.)  When computing the median color profile, to compare with the thick dashed curve in Figure~\ref{fig:gmr1by1}, we weight each galaxy which contributed to a given $R$ bin by the number of spaxels it contributed to the MaNGA-Spaxels measurement.  The thin solid curve shows this profile:  it lies about 0.03~mags below (i.e. blueward of) the profiles from the spectra.  

While 0.03~mags may not seem like much, the scatter in the color magnitude relation is only of order 0.05~mags, so this offset is significant.  If not accounted for, it may lead to biases in SSP analyses of the full spectrum shape.  (A redward tilt will tend to overestimate the ages and or [M/H].  E.g., 0.03~mags could change [M/H] by more than 0.1~dex.)  Working with Lick indices instead avoids such concerns.

However, absorption line work requires high S/N spectra, which we achieve by stacking.  This raises the question of how the color obtained from a stacked spectrum compares with that from the photometry and from the spaxels themselves.  This is non-trivial because absorption line stacks are made by first shifting all spaxels to restframe.  In addition, it is not obvious that the median$(g-r)$ value will be the same as $g-r$ measured in the median spectrum (they are the same for the mean, of course).

To test if this matters, we normalize each (restframe-shifted) spaxel to have the same flux at 6780--6867 \AA\ (while some normalization is necessary, the results to follow do not depend on this particular choice). We then use these normalized spaxels to define a median restframe spectrum.  Finally, we shift each median spectrum by the median redshift of the spaxels in the stack to mimic a median observed spectrum, from which we estimate $g-r$ just as we did for the individual spaxels.  This yields the black solid curve (MaNGA-Stacks) in Figure~\ref{fig:gmr1by1}.  It is in quite good agreement with the thick dashed curve (MaNGA-Spaxels) suggesting that working with quantities measured in the median stack does not lead to a bias.  However, since both MaNGA-Spaxels and MaNGA-Stacks are similarly offset from the color gradients in the photometry, we believe it is prudent to {\em not} fit SSP models to the full spectrum.\footnote{When fitting the full spectrum, some authors \cite[e.g.][]{Vaughan2018a} include a high order polynomial to account for mismatch between the models and the observed spectrum.  While this polynomial might also absorb some of the broadband issues which lead to the discrepancy between photometric and spectro-photometric color, it is not obvious that treating data-data systematics as being on the same footing as data-model issues is justified.  This not an issue for Lick-index based analyses.}
Instead, we now turn to a study of SSP estimates which are based on absorption line strengths in stacked spectra.

\subsection{Normalization when stacking}\label{sec:norm}
Henceforth, we always work with spectra that have been shifted to restframe.

One must decide how to normalize each (restframe) spectrum prior to stacking. One could normalize by the flux around 6800\AA\ (just as we did in our analysis of the $g-r$ color) -- a `global' normalization.  However, the philosophy which underlies Lick index analyses is that broad band features across the spectrum should be removed from the analysis \citep{Worthey1994, Trager1998}.  Therefore, each Lick index includes a wavelength range called the `pseudo-continuum' which is used to normalize the spectrum, as this `local' normalization greatly mitigates problems arising from extinction and flux calibration. This raises the question of whether one should first do a global normalization, then stack, then measure the pseudo-continuum and hence index strength from the single stacked spectrum, or if instead, one should use the pseudo-continuum to normalize each spectrum {\em prior} to including it in the stack, so that the stack which results is different for each Lick index.

\begin{figure}
    \centering
    \includegraphics[width=0.99\linewidth]{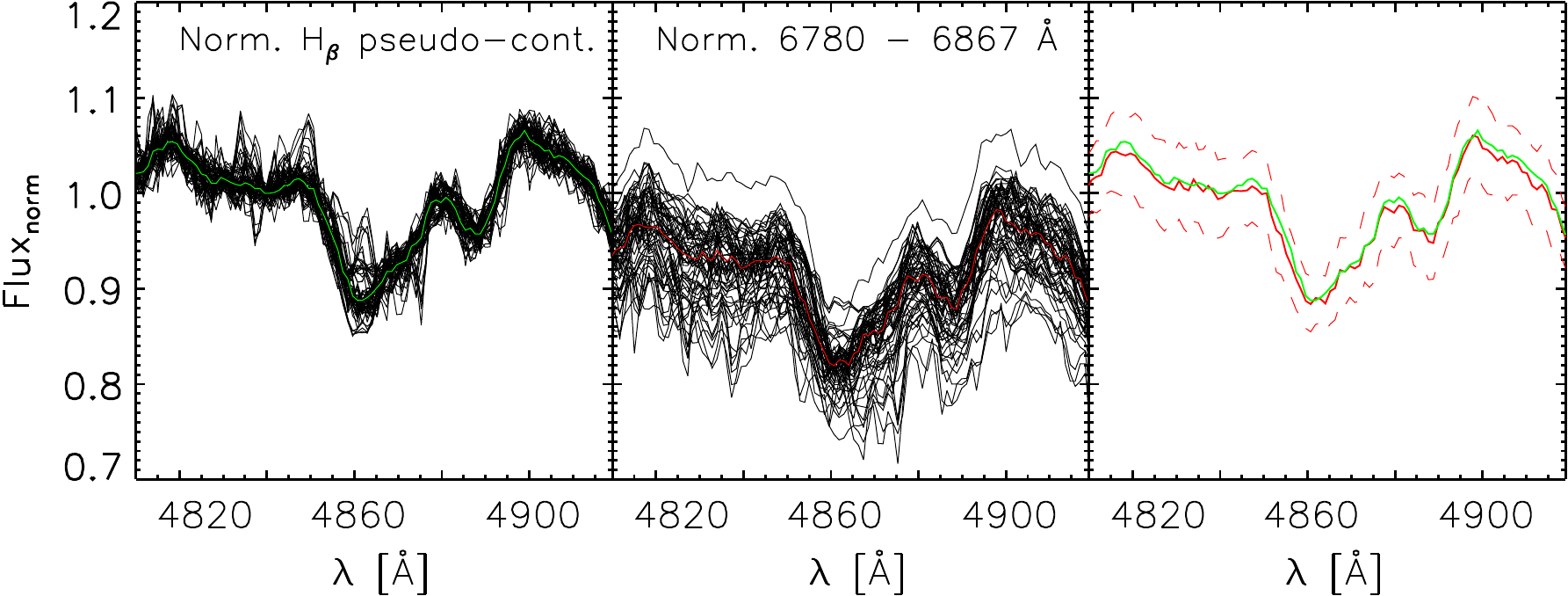}
    \includegraphics[width=0.99\linewidth]{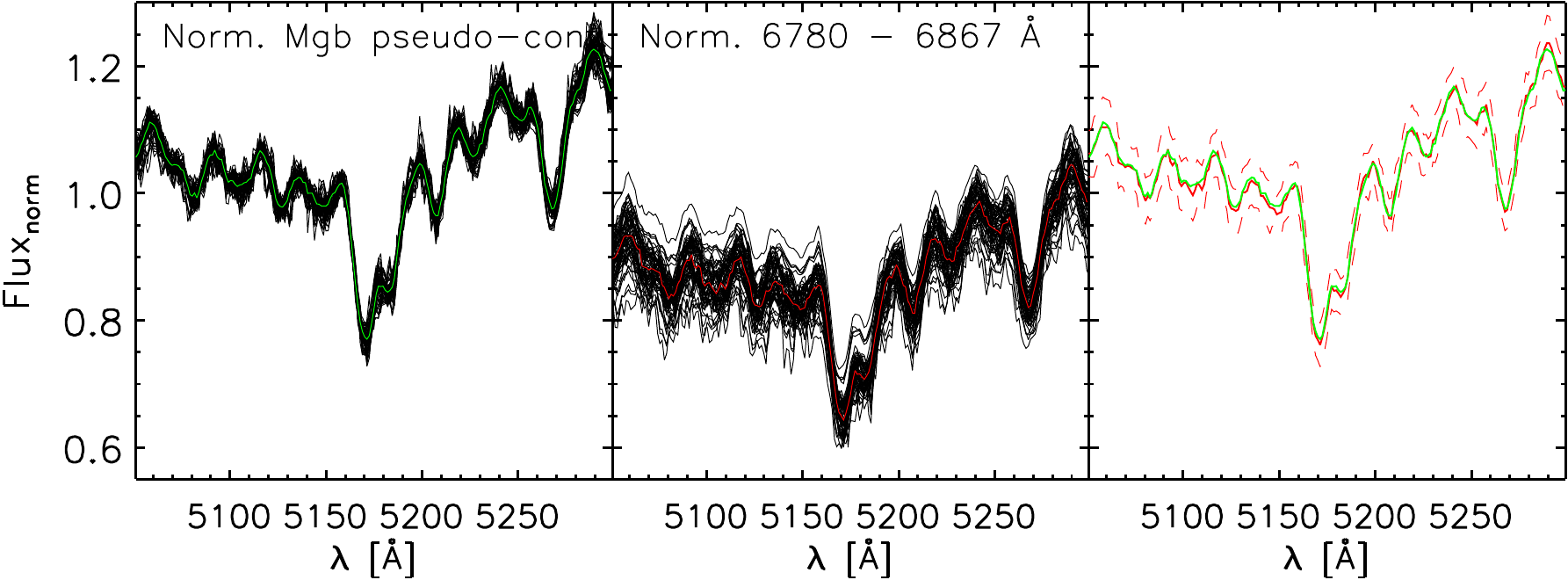}
    \caption{Importance of normalization when measuring the H$_\beta$ (top) and Mg$b$ (bottom) Lick indices from stacked spectra. Black lines show individual spectra (spaxels) of the central region of one of the bins used in this work. Solid green and red lines show the median spectrum when each spaxel is normalized by its pseudo-continuum prior to stacking (left panel -- less noisy) and when the normalization uses the flux in a more distant part of the spectrum (middle panel -- more noisy). The two median shapes agree if each is normalized by its pseudo-continuum (compare red and green solid lines in the right panel) but the rms scatter (dashed lines) is about $3\times$ larger than for our stacking procedure (not shown for clarity, but compare scatter in left panel to that in right).}
    \label{fig:normHb}
\end{figure}

Figure~\ref{fig:normHb} compares these two approaches, which we will refer to as applying `global' vs `local' normalization prior to stacking.  The left hand panel shows the local approach; the middle panel shows the global. In both panels, the bold curve shows the median spectrum which results.  The result of normalizing each median curve using the (index dependent) pseudo-continuum is shown in the panel on the right.  This makes little difference to the curve from the left hand panel (the local approach), of course, but is necessary for the curve from the middle panel (the global approach).  While the two curves are in good agreement, the {\em scatter} around the median is significantly smaller for the local approach: the red dashed lines show the band which includes 68\% of the objects centered around the median for the middle panel (the global approach); although we do not show it (for clarity), the corresponding band for the local approach is about $3\times$ smaller. This is why we believe the local approach is better, and we use it in the remainder of this paper.

If one thinks of normalization as a weight applied to each spectrum, then the stack is a weighted average over the population.  However, because this `weight' is index-dependent, subsequent analysis of Lick indices in `locally' stacked spectra is {\em fundamentally} different from Lick analyses based on a `globally' stacked spectrum, because the latter will typically not be built from the same wavelength-dependent weights (typically, the weights will be based on the signal-to-noise ratio). Stated differently, it is better to stack locally normalized Lick indices than to measure Lick indices in a stack in which spectra were globally normalized.
This is an important difference between our approach and the Lick index-based analyses of \cite{Conroy2014}, \cite{Greene2015} and \cite{Parikh2019}.  While this makes our (local) approach obviously different from approaches which seek to fit the full spectrum \citep{Conroy2018, FeldmeierKrause2021}, \cite{Choi2019} acknowledge that ``the continuum information is largely contained in the narrow absorption features''.  We show this explicitly in Section~\ref{sec:colors}, when we compare measured and predicted broad-band colors.  

\begin{figure}
  \centering
  \includegraphics[width=\linewidth]{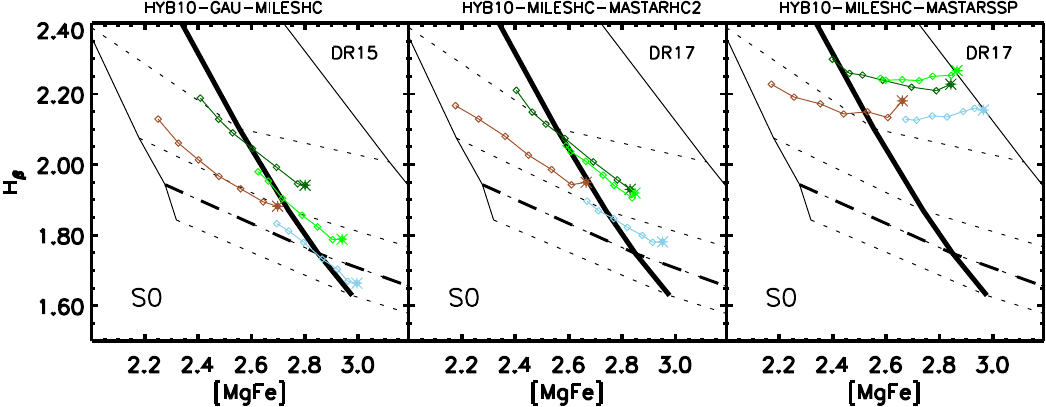}
   \caption{Effect of three different emission corrections on the H$_\beta$-[MgFe] diagnostic plot, for S0s in four $L$ and $\sigma_0$ bins (same in each panel).  Symbols show measurements (smoothed to an effective velocity dispersion of 300~km~s$^{-1}$) as one moves from the center (large asterisk) outwards. Grids, same in each panel, show age and total metallicity [M/H] for solar abundances and a Kroupa IMF. Thick solid line shows solar metallicity and thick dot-dashed line shows an age of 10~Gyrs; younger ages have larger H$_\beta$, and larger [M/H] has larger [MgFe]. }
  \label{fig:HbMgFeS0}
\end{figure}

\subsection{Correction for emission}\label{sec:emission}
Simple stellar population models indicate that the H$_\beta$-[MgFe] index combination nicely separates out age from metallicity \citep{GonzalezPhD, Worthey1994} and are not sensitive to $\alpha$-element abundances \citep{GonzalezPhD, Thomas2005}.  However, the H$_\beta$ absorption line at 4862\AA\ can be filled-in by emission from nearby gas \cite[e.g. see discussion in][]{Poole2010}.  If not corrected-for, such emission will lead to biases in inferred stellar population parameters.  Where emission is obvious, the equivalent width of H$_\beta$ emission is correlated with that of H$_\alpha$ (central wavelength of 6565\AA), which is typically easier to measure.  The correlation is such that the ratio of H$_\alpha$ to H$_\beta$ emission EWs is typically {\em greater} than 2.85 \citep{OsterbrockFerland2006}.  
(Although emission can also be inferred from the [NII] emission lines, at central wavelengths at 6549\AA\ and 6585\AA, they are not tightly correlated with H$_\alpha$ or H$_\beta$.)  But when emission in H$_\beta$ is less obvious, correcting for it is more difficult. 

In MaNGA, the emission is estimated by fitting simple stellar population spectra, spanning a coarse grid in age and metallicity and with IMF fixed to Salpeter, to the observed spectrum.  Whereas MaNGA DR15 provided a single estimate for the strength of the H$_\beta$ emission in each spaxel, MaNGA DR17 provides two.  The two DR17 values use different continuum templates (known as MASTARSSP vs MASTARHC) in the emission-line fitting module \cite[for details about this, as well as other more minor changes to the reduction pipeline, see][and {\tt www.sdss.org/dr17/manga/manga-data/whatsnew}]{Law2021}.

Figure~\ref{fig:HbMgFeS0} shows the result of using the three different emission corrections and plotting the H$_\beta$ absorption line strength versus [MgFe] measured in a number of stacked spectra.  (No emission correction is needed for the [MgFe] line.)  Following common practice, all spectra were first smoothed to an effective velocity dispersion of 300~km/s.  The four sets of symbols in each panel represent S0s in four bins of $L$ and $\sigma_0$; the large asterisk shows the central region and the other symbols show the values at larger radii.  We have chosen to only show the S0s  -- for which H$_\beta$ emission is usually obvious -- and, to isolate the changes arising from the corrections, rather than the sample itself, we only use the S0s which were present in DR15.  (They account for about half of the full DR17 S0 sample which we study in Sections~\ref{sec:models} and~\ref{sec:miles}.)

\begin{figure}
  \centering
  \includegraphics[width=0.99\linewidth]{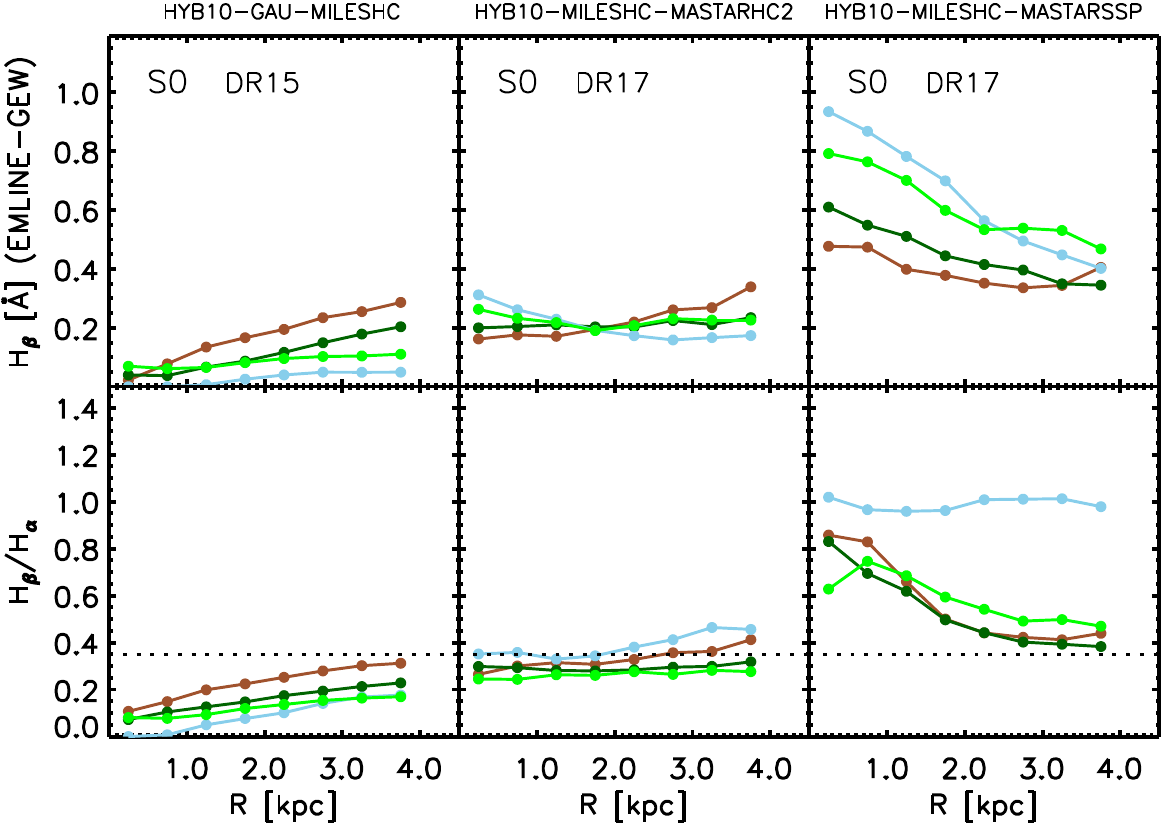}
  \caption{Comparison of H$_\beta$ (top) and H$_\beta$/H$_\alpha$ (bottom) emission line strengths in MaNGA DR15 (left) and DR17 (middle and right), for S0s in the same four bins as in the previous figure.   Dotted line in each bottom panel shows 0.35 (=1/2.85), the approximate expected lower limit.  Ratios shown in the bottom panels suggest that the DR15 values imply unrealistic amounts of extinction in the central regions, and that of the two DR17 estimates, those in the middle panel are more likely to be correct.}
  \label{fig:oldS0}
\end{figure}

\begin{figure}
  \centering
  \includegraphics[width=0.89\linewidth]{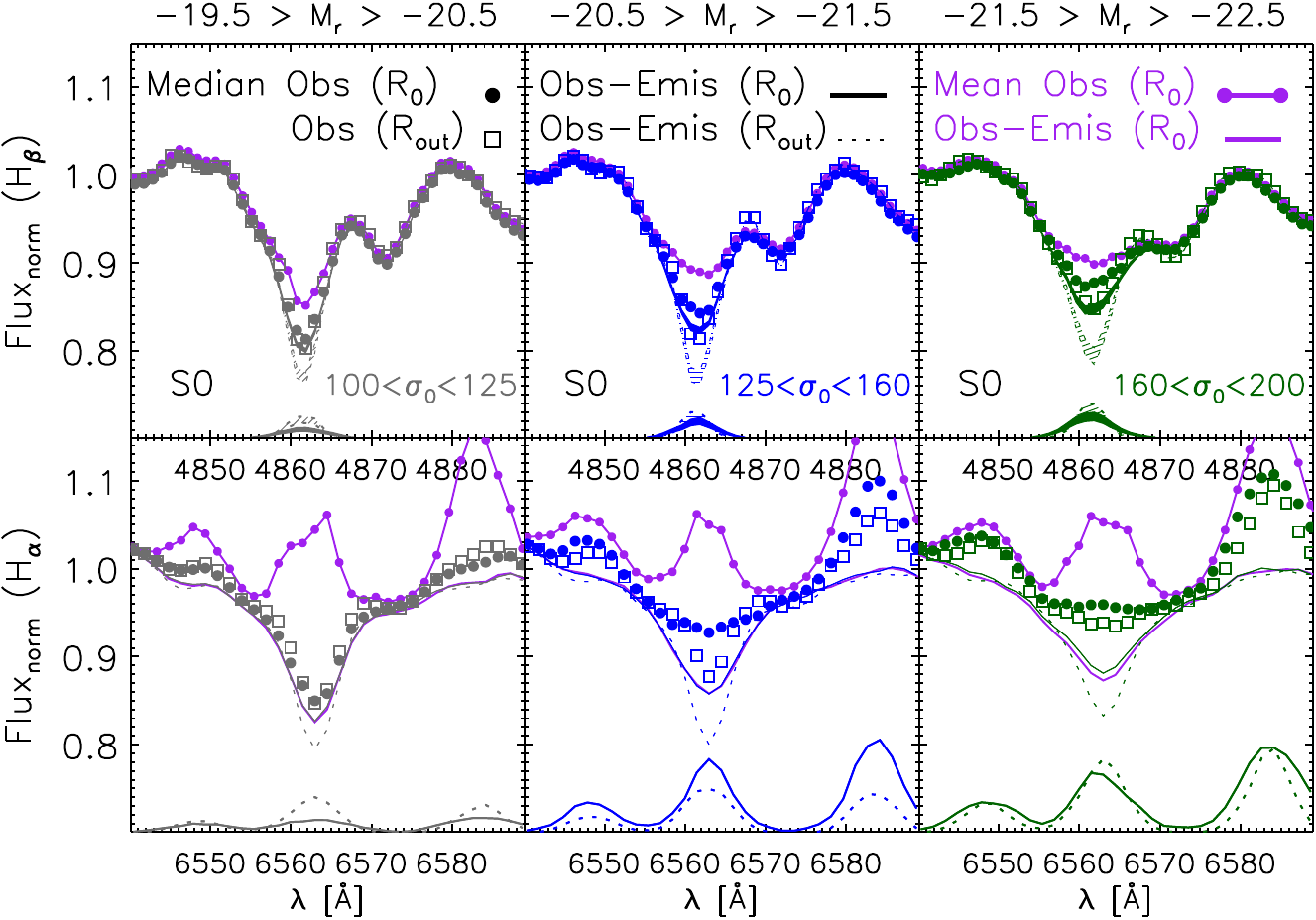}
  \caption{Observed (symbols) and emission-corrected (curves) H$_\beta$ (top) and H$_\alpha$ (bottom) {\it median} line profile shapes in the central region (filled circles and solid curve) and further out (typically about $R_e$, open squares and dotted curve). Purple filled circles and solid lines show the {\it mean} profiles in the central region (for comparison with the {\it median} values). Panels show results for S0s in three bins in luminosity and $\sigma_0$, and curves along the bottom of each panel show the {\it median} emission shifted upwards by 0.65.  In top panels, hashed regions show an estimate of the systematic uncertainty in the line shape which is discussed in the text.
  }
  \label{fig:HbHaLine}
\end{figure}

Superposed in each panel are (the same) model age vs. total metallicity [M/H] grids (for solar abundances and a Kroupa IMF).  These serve to guide the eye, and to show that the DR15 corrections (left) result in older less metal rich SSPs than the DR17 MASTARSSP corrections (right); the implied gradients are also rather different.  The DR15 corrections were used by \cite{DS2019}, so we expect some of their results to be modified if we use either of the newer DR17 corrections.

To study the relative merits of the three corrections, Figure~\ref{fig:oldS0} compares the H$_\beta$ and H$_\beta$/H$_\alpha$ emission line-strengths for the same set of S0s as in the previous figure.  The four sets of curves in each panel show average emission line-strength profiles for S0s of different luminosity.  We noted earlier that we expect H$_\alpha$/H$_\beta$ to be of order 3 or larger, so we believe the H$_\beta$/H$_\alpha\sim 1$ values in the bottom right panel are unrealistic, despite the fact that they are recommended by the MaNGA collaboration.  Although H$_\beta$/H$_\alpha < 0.35$ values could be indicative of dust, ratios as small as seen at small $R$ in the bottom left panel would require substantial amounts of dust in the central regions, suggesting that the DR15 values were incorrect.  The middle panel has H$_\alpha$/H$_\beta\sim 2.8$, which is close to the expected ratio, so the {\tt HYB10-MILESHC-MASTARHC2} reductions are our fiducial choice.  It is these values which we adopt as our standard in the remainder of this paper, as well as in \cite{Bernardi2022}.  Since the DR15 emission corrections were different from our fiducial choice, our SSP parameters (and gradients) will differ from those of \cite{DS2019}.  

\begin{figure}
  \centering
  \includegraphics[width=0.89\linewidth]{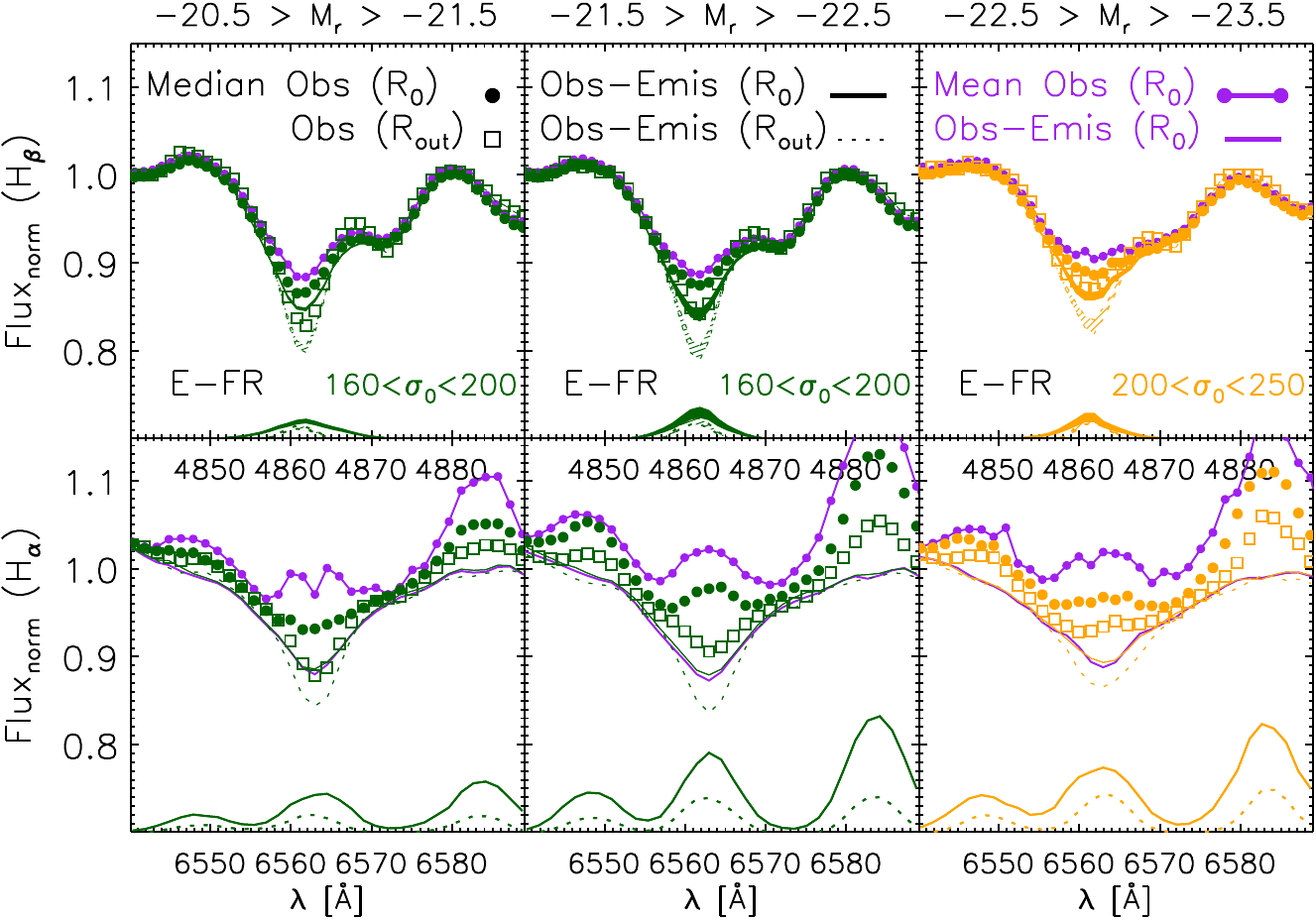}
  \includegraphics[width=0.89\linewidth]{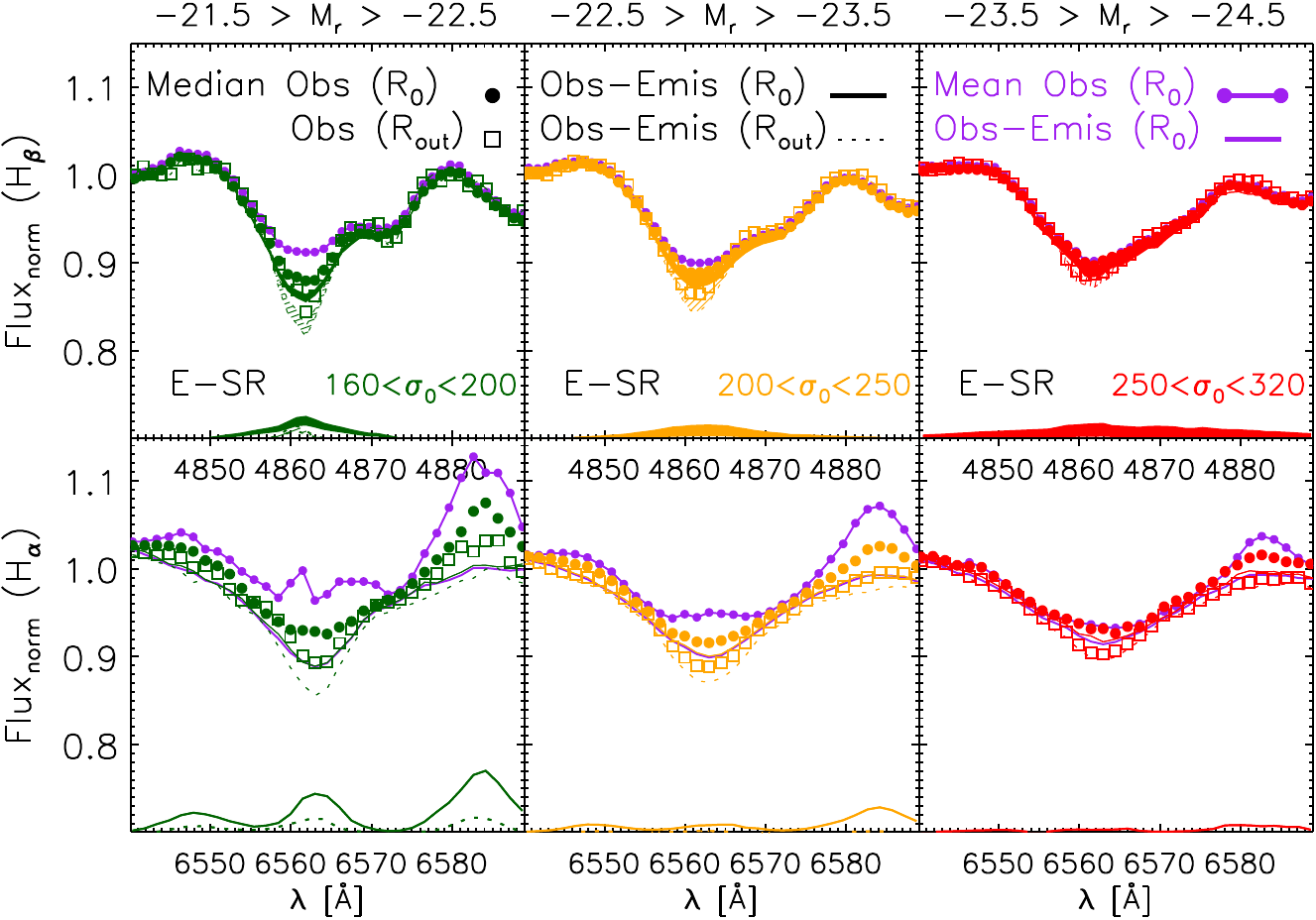}
  \caption{Same as Figure~\ref{fig:HbHaLine}, but now for Ellipticals that are fast- or slow-rotators (top and bottom, respectively).  Fill-in by emission is much less obvious for E-SRs, and the emission in the central spaxel is stronger than at larger $R$.  }
  \label{fig:HbHaESR}
\end{figure}

\subsection{Systematic uncertainties on emission correction}\label{sec:emierr}
Unfortunately, the MaNGA database does not provide an estimate of the uncertainty on the emission correction.  Therefore, we estimate uncertainties as follows.

To illustrate, Figure~\ref{fig:HbHaLine} shows the H$_\alpha$ (bottom) and H$_\beta$ (top) median line profiles in stacked S0 spectra, before (symbols) and after (curves) subtracting the fiducial emission correction.  (We describe the hashed regions in the top panels shortly.)  Each panel shows the median stacked spectrum for objects in a narrow bin in $L$ and $\sigma_0$ as labeled.  (This and the next figure are the only ones where we have not smoothed the spectra to an effective resolution of 300~km/s.)
The two sets of symbols and curves in each panel show the spectrum in the central regions and in an outer shell whose scale depends on $L$ and $\sigma_0$ (it is the largest $R_{\rm maj}$ shown in Figure~\ref{fig:miles}, and is typically a little smaller than $R_e$).  
The purple curves show the mean, rather than median, of the central stack, prior to and after correcting for emission.  Whereas the mean is very different from the median prior to correction, because the distribution of emission line strengths is very skewed, it is almost indistinguishable from the median after correction, suggesting that the emission correction has removed the most dramatic variations.  (This is also true for non-central regions.)  The curves along the bottom of each panel show the median emission.  It differs from the median of the difference between the observed and corrected spectra and from the difference of the median values, neither of which are shown, because the emission strength has such a skewed distribution.

For H$_\alpha$ (bottom panels), the presence of emission is obvious even by eye.  After accounting for emission, the continuum shape far from the center of the H$_\alpha$ line is similar; differences remain only in the line centers.  The presence of emission in H$_\alpha$ suggests it must also be accounted for in H$_\beta$.  However, the top panels show that, even for S0s, it is hard to `see' this emission in the H$_\beta$ line shapes.

We estimate the uncertainty on the H$_\beta$ absorption line by making use of the limiting value of 2.85 for the ratio of H$_\alpha$ to H$_\beta$ emission EWs (c.f. Figure~\ref{fig:oldS0} and related discussion). Firstly, we assume a 10\% uncertainty $\epsilon$ on the H$_\alpha$ and H$_\beta$ EWs.  The values H$_\alpha \pm \epsilon_\alpha$ and H$_\beta\pm\epsilon_\beta$ yield four values for the ratio of H$_\alpha$ to H$_\beta$ emission EWs.  For each such ratio:  if it is smaller than 2.85, then we assume that H$_\beta$ has been overestimated, so we reduce its amplitude (at all wavelengths) by the amount needed to make this ratio equal 2.85 (e.g., if the ratio is 2.85/2, then we set H$_\beta\to$H$_\beta/2$ at all wavelengths). If the ratio is less than 0.2, we set it equal to 0.2 and we then decrease H$_\beta$ by the amount needed to make the ratio equal 2.85.  If the ratio exceeds 2.85, we do not apply any further correction (i.e. the uncertainty is 10\%).  This is because the ratio of emission EWs could exceed 2.85 because of extinction (this is known as the Balmer decrement).  The hashed and shaded regions in the top panels show the error bands on the emission corrected H$_\beta$ absorption line which result.

Correcting for emission is even more difficult for E-FRs and E-SRs, which have larger $\sigma$ {\em and} smaller emission (see Figure~\ref{fig:HbHaESR}).  For these samples, comparison with a model is important for determining the emission correction to H$_\beta$.  Notice that the central regions of both E-FRs and E-SRs have stronger emission than at the larger $R$.  In addition, although the emission corrected profiles tend to be deeper at the largest $R$ considered here, some of this is simply a consequence of the velocity dispersion being larger in the central regions; the actual EW is not so different.  Therefore, if we find slightly younger stellar populations in the central regions of E-FR and E-SRs, we should not be too surprised.

\section{Models}\label{sec:models}
As we noted in the Introduction, we determine light-weighted\footnote{Light-weighted properties are sensitive to the presence of younger stars, owing to the fact that young stars outshine older stars.} stellar population parameters by fitting SSPs to the measured line strengths.  This raises the question of which indices to use, and which models to fit.  For reasons that will become clear, our fiducial model is MILES+Padova.  This model characterizes an SSP by its age, total metallicity [M/H], $\alpha$-element abundance [$\alpha$/Fe], and IMF.\footnote{Strictly speaking, the Padova isochrones are not $\alpha$-enhanced, so, as described in \cite{DS2019}, we calibrate responses using the BaSTi isochrones, which we then apply to MILES+Padova.}

For the MILES models, $\alpha$ is a combination of O, Ne, Mg, Si, S, Ca and Ti.  We use [Z/H] to denote the metallicity assuming solar element abundance patterns.  It can differ from the total metallicity [M/H] if element abundance ratios differ from solar.  Note that [Z/H] is often denoted [Fe/H], but, for C18, [Fe/H] denotes Iron abundance alone.  Thus, 
\begin{equation}
    \left[\frac{\rm M}{H}\right] = 
      \left[\frac{\rm Fe}{H}\right] + \log_{10}\left(Z_{0} + Z_{\rm C}\,10^{[{\rm C}/{\rm Fe}]} + Z_\alpha\, 10^{[\alpha/{\rm Fe}]}\right) , 
    \label{eq:MH}
\end{equation}
where $Z_{\rm C}$ is the fraction of the Sun's mass in metals that is in C,
$Z_\alpha$ is the fraction that is in $\alpha$-elements, 
and $Z_0 = 1 - Z_{\rm C} - Z_\alpha$.
The MILES models use the solar abundances given in \cite{Sun1998}.  Converting the number weighted abundances in their Table~1 to mass-weighted abundances yields    $(Z_{\rm C},Z_\alpha) = (0.133,0.694)$.

Strictly speaking, the MILES models have [C/Fe]=0, so the expression above is usually written in terms of $Z_0+Z_{\rm C}=1-Z_\alpha$ and $Z_\alpha$ \cite[following, e.g.,][]{Salaris1993}.  We have used non-zero [C/Fe] to show how this expression is modified when other elements are individually enhanced.  (E.g., if C and Na are each enhanced, one would add $Z_{\rm Na}\,10^{[{\rm Na}/{\rm Fe}]}$ to the terms in the big round brackets, and set $Z_0 = 1 - Z_{\rm Na} - Z_{\rm C} - Z_\alpha$.)

The MILES model grids are phrased in terms of constant [M/H] (also see \citealt{Trager2000} and \citealt{TMB2003}), so variations in [$\alpha$/Fe] imply variations in [Z/H].  In contrast, the C18 models allow one to enhance some elements individually while keeping [Z/H] fixed.  For these models, equation~(\ref{eq:MH}) (and how it generalizes) is particularly useful, since it allows one to track how [M/H] changes as, e.g., [$\alpha$/Fe] is varied.  However, note that the C18 models use the solar abundances of \cite{Sun2009}, for which the coefficients in equation~(\ref{eq:MH}) are modified to
  $(Z_{\rm C},Z_\alpha) = (0.180,0.663)$.
Enhancing elements individually provides intuition into how different line index strengths respond to changes in enhancements.  We explore this in some detail in Appendix~A, where we show why, for the MILES models, the line-index combinations which provide constraints that are relatively straightforward to visualize are:\\
\indent\indent H$_\beta$-[MgFe], $\langle {\rm Fe}\rangle$-[MgFe] and TiO$_{\rm 2SDSS}$-[MgFe].  \\
These indices have previously been shown to be useful by, e.g. \cite{GonzalezPhD}, \cite{Worthey1994}, \cite{Trager1998}, \cite{Thomas2005} and \cite{LaBarbera2013}, and we will use these combinations extensively in what follows.
Appendix~A also describes how we build [$\alpha$/Fe] enhanced C18 models.

It is important to note that the discussion below is intended to highlight which conclusions about early-type galaxy stellar populations are common to MILES and C18 and which are not.  Determining why they sometimes disagree is beyond the scope of our work, but see \cite{Vazdekis2015} and \cite{Conroy2018} for a discussion of some known differences.

\subsection{Parametrization of IMF}
We noted in the Introduction that we are particularly interested in IMF gradients.  
There are a number of parametrizations of the IMF, $dN/dm$, ranging from\\
- a single power law (called `unimodal' in MILES); if the slope equals $-2.35$ (MILES $\Gamma=1.3$), this is known as a \cite{Salpeter1955} IMF; \\
- a power-law which is truncated at lower masses in a way which depends on the power law slope (called `bimodal' in MILES; an IMF with $\Gamma_b=1.3$ is very close to that of \citealt{Kroupa2001}), and \\
- a power law whose slope is fixed to $-2.35$ at high masses, but transitions to something flatter at lower masses (popular examples are the Kroupa and Chabrier IMFs).  
In the C18 models, this transition is controlled by two parameters, $\alpha_1$ and $\alpha_2$, the slopes at low masses, such that 
 $dN/dm$ is proportional to $m^{-\alpha_1}$ for  $0.08 < M/M_\odot \le 0.5$, 
 to $m^{-\alpha_2}$ for  $0.5< M/M_\odot \le 1$, 
 and to $m^{-2.35}$ for $M\ge M_\odot$.  

\begin{figure}
    \centering
    \includegraphics[width=0.85\linewidth]{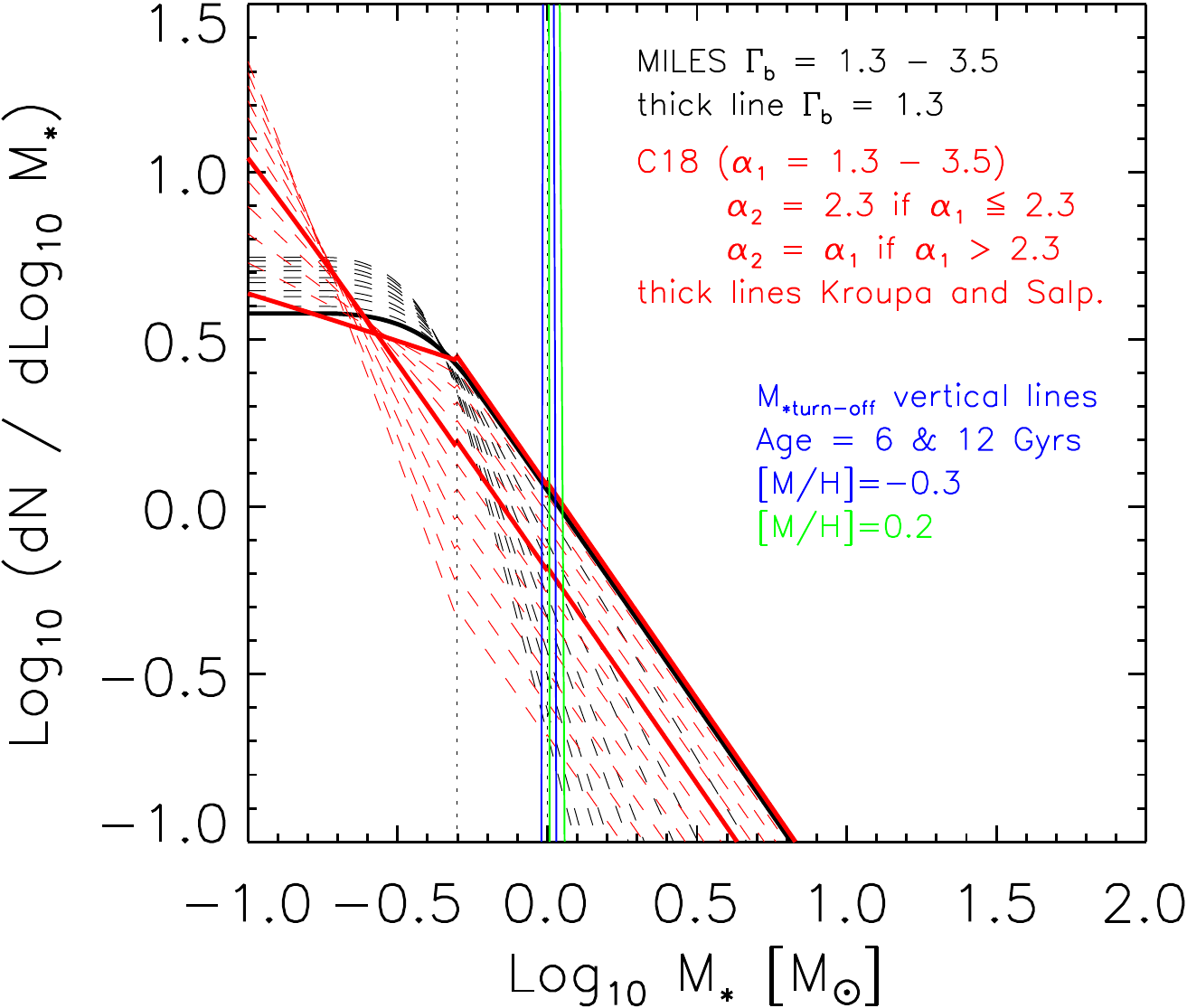}
    \caption{Range of IMFs for the MILES (black) and C18 (red) models we consider in our study.  Thick curves show the Kroupa and Salpeter IMFs for comparison.  Dotted vertical lines show the mass scales at which the C18 IMFs change slope.  Solid vertical lines show the main sequence turnoff, which depends on age and [M/H]:  stars more massive than this contribute to the mass in remnants and in gas.   The $M_*$ values we quote include the mass in stars still on the main sequence (i.e. less massive than the turnoff mass) and the mass in remnants, but not the mass in gas.}
    \label{fig:imfs}
\end{figure}

 The shape of the MILES Bimodal IMF is controlled by a single parameter, $\Gamma_b$. As noted above, when $\Gamma_b=1.3$, the IMF is very close to that of \cite{Kroupa2001}.  Increasing $\Gamma_b$ steepens the slope at high masses, thus increasing the relative contribution from lower mass stars.  So as to work with only one free parameter for the C18 models as well, we set $\alpha_2=2.3$ and let $\alpha_1$ vary from $0.9 - 3.5$.  This ensures that $\alpha=1.3$ and $\alpha=2.3$ yield a shape that is close to Kroupa and Salpeter, respectively.  Increasing $\alpha_1$ increases the abundance of low mass stars (i.e. like increasing $\Gamma_b$).  In addition, if $\alpha_1 > 2.3$, we also consider models having $\alpha_2=\alpha_1$ (rather than fixing $\alpha_2=2.3$); this further increases the abundance of low mass stars.  However, in contrast to MILES, the C18 models fix the high mass end to Salpeter, and only modify the abundances at lower masses.  Figure~\ref{fig:imfs} shows these differences.

\begin{figure}
    \centering
    \includegraphics[width=0.85\linewidth]{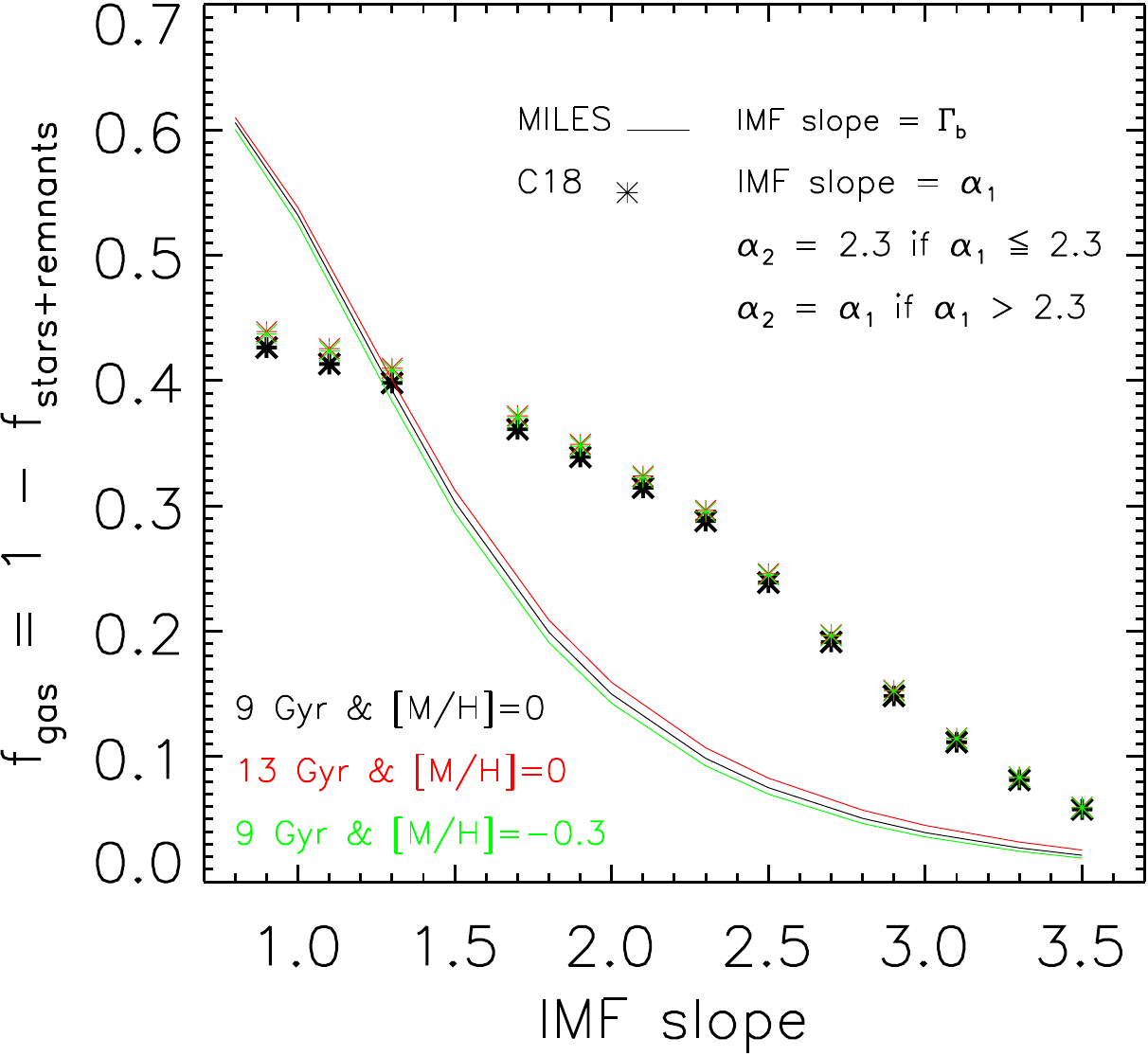}
    \caption{Gas fraction depends strongly on IMF shape: bottom-heavy IMFs have smaller gas fractions, but $f_{\rm gas}$ depends strongly on how the IMF was parameterized.  Increasing $\Gamma_b$ decreases the abundance of massive stars and hence $f_{\rm gas}$; increasing $\alpha_1$ does not modify the high-mass slope, so the effect on $f_{\rm gas}$ is less dramatic.  For sufficiently old stellar populations, the dependence on $\Gamma_b$ and $\alpha_1$ is approximately independent of age and [M/H]. 
    }
    \label{fig:fgas}
\end{figure}

In Appendix~A, we study the response of various indices to changes in $\Gamma_b$ and $\alpha_1$, which then impact the determination of SSP parameters from observed index strengths.  While the difference in IMF parametrization can result in subtle systematic effects on the inferred values of SSP parameters, differences in the predicted gas mass fractions -- which arise as stars evolve off the main sequence -- are more dramatic.  Figure~\ref{fig:fgas} shows that an SSP that is more than a few Gyrs old and formed with a Kroupa-like IMF has a gas mass fraction of $f_{\rm gas}\approx 0.4$ (for both MILES and C18); in contrast, an IMF with $\Gamma_b=2.3$ has $f_{\rm gas}\sim 0.1$, whereas an IMF with $\alpha_1=2.3$ has $f_{\rm gas}\sim 0.3$.  Exploiting $f_{\rm gas}$ as an IMF-indicator is beyond the scope of this work.

\subsection{Choice of model grids}
\begin{figure*}
  \centering
  \includegraphics[width=0.85\linewidth]{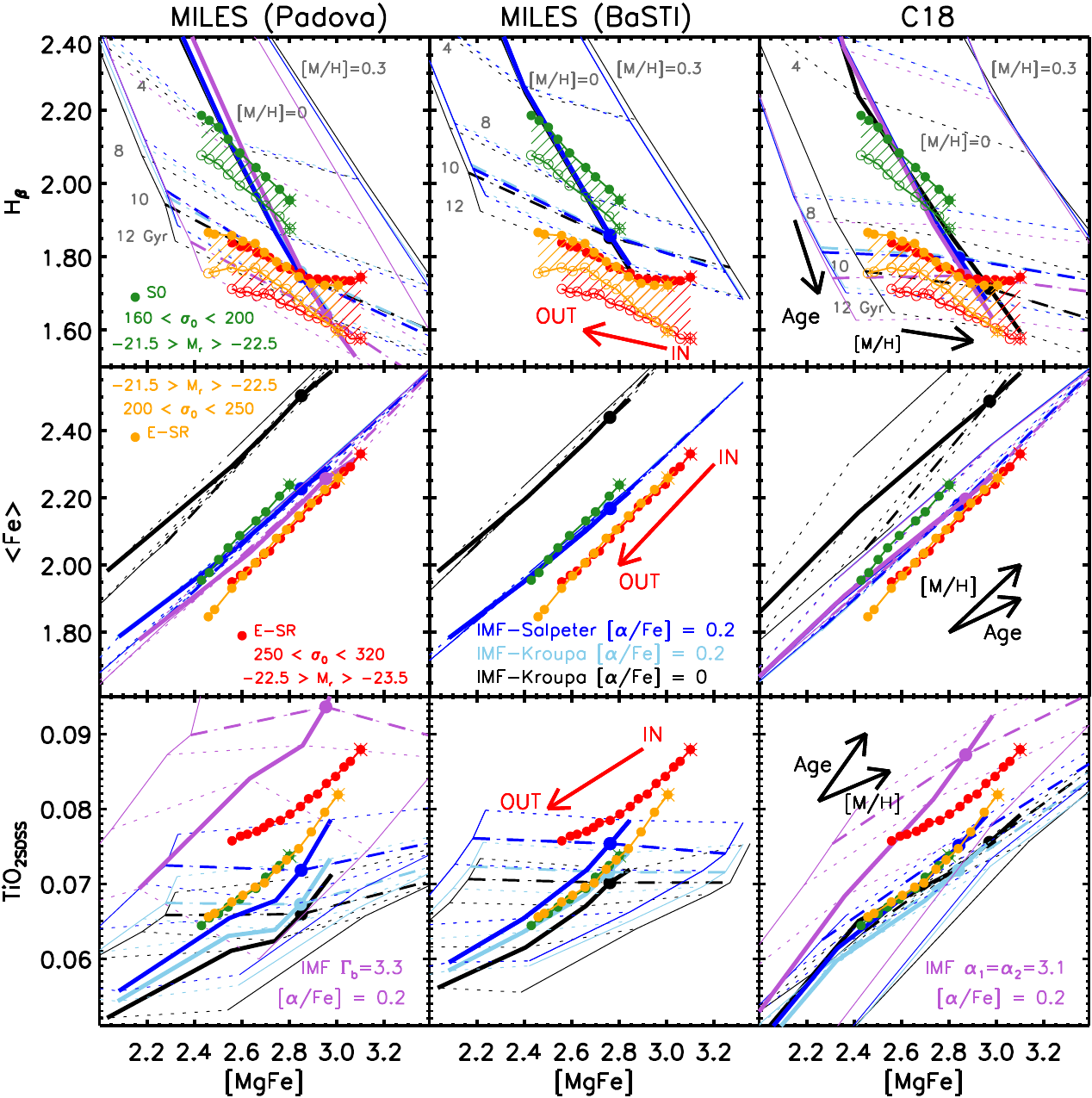}  
  \caption{Comparison of MILES (left and middle) and C18 (right) model age vs total metallicity [M/H] grids, for a few representative values of [$\alpha$/Fe] and IMF shape (as labeled),  with line strengths (filled symbols) measured from stacked spectra of objects in various luminosity and velocity dispersion bins (as labeled).  Asterisk shows the central value, and the others show progressively larger scales.  Solid lines show loci of fixed [M/H] (thick solid line is solar), and dotted lines show loci of fixed age (thick dashed line shows an age of 10~Gyrs).  Statistical errors on the measurements are smaller than the symbol sizes.  Filled and open symbols in top panels were obtained from the larger and smaller emission corrections to H$_\beta$ discussed in the text (e.g. Figure~\ref{fig:HbHaESR}); the (shaded) region between the two serves as a crude estimate of the systematic error on H$_\beta$.
  Stellar population parameters, age, [M/H], [$\alpha$/Fe] and IMF are determined by finding those values which simultaneously reproduce the observed H$_\beta$, $\langle$Fe$\rangle$, TiO$_{\rm 2SDSS}$ and [MgFe] line strengths.  
  }
  \label{fig:compare}
\end{figure*}

In what follows, we will use the line index combinations H$_\beta$-[MgFe], $\langle {\rm Fe}\rangle$-[MgFe] and TiO$_{\rm 2SDSS}$-[MgFe] to constrain four SSP parameters:  age, metallicity, [$\alpha$/Fe] and IMF.  To illustrate the methodology, we show how the different index pairs are expected to depend on age and metallicity when the IMF is fixed to Kroupa and [$\alpha$/Fe] is changed from 0 to 0.2, as well as when the IMF is Salpeter with [$\alpha$/Fe]=0.2 (black, cyan and blue grids in Figure~\ref{fig:compare}).  Finally, purple grids show models with [$\alpha$/Fe]=0.2 and extremely bottom-heavy IMFs.
For now, we are only interested in comparing the model grids with one another; we discuss the measurements, shown as symbols, in the next two subsections.  
The left hand panels show the MILES+Padova models, middle panels show MILES+BaSTI and right hand panels show C18.

The top left panel shows a few model age-[M/H] (i.e. total metallicity) grids in the H$_\beta$-[MgFe] plane, for the MILES+Padova model.  For now, only consider the black grids, which are for a Kroupa IMF (MILES Bimodal $\Gamma_b=1.3$) and solar abundance ratios (i.e., [M/H]=[Fe/H]).  The solid lines show loci of fixed total metallicity [M/H] (thick solid line is solar), and dotted lines show loci of fixed age (thick dashed line shows an age of 10~Gyrs).  The first thing to notice is that, in this plane, age and [M/H] are nicely separated \cite[e.g.][]{Worthey1994}.  Increasing [$\alpha$/Fe] but keeping the IMF fixed to Kroupa makes no difference (cyan grid).  Changing the IMF from Kroupa to Salpeter makes no difference (blue grid).  Only when the IMF is even more bottom heavy (purple grid is for [$\alpha$/Fe] = 0.2 {\em and} $\Gamma_b=3.3$) is there some effect:  increasing $\Gamma_b$ reduces the predicted H$_\beta$ (slightly changing [MgFe]) making the grids slide down and slightly to the right, so that there is a slight degeneracy between the age-[M/H] values and IMF.  Except for this, H$_\beta$-[MgFe] provides a good first guess for the age and [M/H].  

In the second row, the left hand panel shows age-[M/H] grids for the same set of models, but now in the $\langle$Fe$\rangle$-[MgFe] plane.  In this case, for a given [$\alpha$/Fe] and IMF, the grids basically define a line, the locus of which is determined almost entirely by [$\alpha$/Fe] -- the IMF hardly matters.  Thus, $\langle {\rm Fe}\rangle$-[MgFe] provides a clean diagnostic of $\alpha$-element abundances \cite[in agreement with many previous studies, e.g.][]{Trager1998,Thomas2005}.  

The bottom left panel shows age-[M/H] grids in then TiO$_{\rm 2SDSS}$-[MgFe] plane.  As for H$_\beta$-[MgFe], age and [M/H] are approximately orthogonal, but now there is a dramatic dependence on IMF; for a given IMF, the dependence on [$\alpha$/Fe] is weak.  Thus, when used in combination with the other two grids, TiO$_{\rm 2SDSS}$-[MgFe] constrains the IMF.  

The middle panels show a similar set of grids, but for the MILES+BaSTI models.  They show the same qualitative trends as MILES+Padova, except that they are offset to larger H$_\beta$ values.  As a result, the MILES+BaSTi models end up implying significantly older ages than MILES+Padova; for more massive objects, the MILES+BaSTi ages can exceed the age of the universe, which is why, of the two MILES models, we prefer MILES+Padova.  

\subsection{Direct comparison of MILES and C18}
We now turn to a study of how the MILES+Padova and C18 SSPs compare for the Kroupa and Salpeter IMFs.  Unfortunately, as Figure~22 of \cite{Vazdekis2015} suggests, for a Salpeter IMF, the MILES age-[M/H] grids differ systematically from those of C18 even at solar metallicity and abundance.  One might have thought that this would be less of an issue for gradients, but, as we show below, things are more complicated.  

Consider the top right panel of Figure~\ref{fig:compare}.  Like MILES, C18 has [MgFe] increasing as [M/H] increases; however, at ages older than about 8~Gyrs, the MILES models have H$_\beta$ decreasing as [M/H] increases whereas C18 has H$_\beta$ nearly constant.  This subtle difference has a profound consequence.  

To see why, consider the red and yellow filled symbols -- same in each of the top panels -- which show the line index strengths measured in stacked spectra of E-SRs having the luminosity and central velocity dispersion values indicated.  The fact that both sets of symbols approximately overlap is a coincidence; other $L$ and $\sigma_0$ bins have line-strengths that are less well-matched.  (The filled green symbols show results from an S0 stack, with the $L$ and $\sigma_0$ values indicated.) The statistical errors on the measurements are smaller than the symbol sizes. The open circles show the result of using the smaller of the two emission corrections to H$_\beta$ discussed in Section~\ref{sec:emission}.  The hashed region between the two is a crude estimate of the systematic error which arises from uncertainties in the correction for emission, and will propagate to the inferred SSP parameters.  (Figure~\ref{fig:errors} addresses this in more detail.)  

The asterisk shows the index strengths in the central stack and the others show progressively larger scales.  For E-SRs, the MILES Kroupa IMF grid indicates that, except for the central region which is slightly younger, the age is approximately constant for most scales and [M/H] decreases with distance from the center.  For C18, however, the model grids are rotated slightly, so the {\em same} measurements imply an age gradient:  the outer parts are younger (leftwards of the thick [M/H]=0 line).  As we show shortly, this difference in gradients is important when inferring the IMF shape.  

The second thing to notice is the effect of changing the $\alpha$-element abundances (black and cyan grids).  Recall that, for the MILES models, the H$_\beta$-[MgFe] combination is nearly independent of [$\alpha$/Fe].  However, for C18, the model grids shift upwards and to the left, so the ages inferred from this plot increase as [$\alpha$/Fe] increases.  

Thirdly, the blue grids show [$\alpha$/Fe]=0.2 and a Salpeter IMF.  These grids are almost indistinguishable from the cyan ones, both for MILES and for C18, suggesting that the H$_\beta$-[MgFe] diagnostic is independent of IMF.  However, for extremely bottom heavy IMFs (purple grids show $\Gamma_b=3.3$ in the MILES panel, and $\alpha_1=\alpha_2=3.1$ in the C18 panel) the grids are shifted downwards and to the right, so this IMF will return younger ages but similar [M/H] compared to the cyan or blue grids.  This allows us to anticipate one of our results:  although ages and, hence, age gradients, may depend on the assumed IMF -- and these may differ for MILES and C18 -- both models will produce [M/H] gradients that do not depend on the IMF.   

The next set of panels shows $\langle$Fe$\rangle$ versus [MgFe].  The red and yellow sets of symbols are again very similar, and the measurements strongly prefer models with [$\alpha$/Fe]$ > 0$.  The C18 models will tend to return similar or slightly smaller [$\alpha$/Fe] values than MILES.  While this combination is an excellent diagnostic of [$\alpha$/Fe] regardless of IMF for MILES (left panels), the C18 model grids (right panel) for a given [$\alpha$/Fe] open out to cover a larger part of the plane, so the [$\alpha$/Fe] determination is not as straightforward.  Nevertheless, we expect that for C18 also, [$\alpha$/Fe] gradients will not depend strongly on the assumed IMF.  This fanning out, and the shifting of the H$_\beta$-[MgFe] grids in response to [$\alpha$/Fe], are a second set of important differences between the MILES and C18 models.  

Finally, the bottom panels show TiO$_{\rm 2SDSS}$ vs [MgFe].  Here, the yellow and red sets of symbols define rather different loci; by coincidence, now it is the yellow and green symbols which show substantial overlap.  This illustrates the sense in which TiO$_{\rm 2SDSS}$ provides new information that is `orthogonal' to that provided by H$_\beta$ and $\langle$Fe$\rangle$.  This new information is regarding the IMF:  the grids for different IMFs are clearly offset from one another.  Whereas both MILES and C18 agree that more bottom heavy IMFs predict larger TiO$_{\rm 2SDSS}$ -- and this is particularly true for large $\Gamma_b$ or $\alpha_1$ -- they are otherwise rather different.  

Consider the predicted line strengths for an age of 10~Gyrs and [M/H]=0.  The MILES grids shift slightly to larger TiO$_{\rm 2SDSS}$ when [$\alpha$/Fe] increases, whereas C18 shows the opposite (also compare grey and cyan lines in Figure~\ref{fig:response}).  In addition, for MILES, lines of constant age are almost horizontal (e.g. thick dashed lines): large changes in [M/H] produce little change in TiO$_{\rm 2SDSS}$.  In contrast, for C18, TiO$_{\rm 2SDSS}$ increases as [M/H] increases.  As a result, even for a Kroupa IMF, age and [M/H] are almost completely degenerate for C18, and almost orthogonal for MILES.  Whereas the C18 grids open up as $\alpha_1$ increases, they always have TiO$_{\rm 2SDSS}$ increasing more strongly with [M/H] than MILES.  This third significant difference between the models has the following important consequence for gradients.  

The bottom left and right panels show that, for a given IMF, there is a large age gradient, both for MILES and for C18.  However, recall from the top left panel that the MILES models require approximately constant age (in the outer parts, where [M/H]$<0$) if the IMF is fixed.  Thus, the only way for the MILES models to reconcile the H$_\beta$-[MgFe] gradients with the TiO$_{\rm 2SDSS}$-[MgFe] gradients is by invoking a large IMF gradient.  Indeed, for MILES, the strength of TiO$_{\rm 2SDSS}$ in the central regions of all galaxies can only be explained by rather bottom heavy IMFs.  For C18, the measured H$_\beta$-[MgFe] gradients imply age gradients (at fixed IMF), so the need for IMF gradients is reduced.  

To summarize:  The panels on the left of Figure~\ref{fig:compare} demonstrate that, for the MILES models:\\
 \indent H$_\beta$ vs [MgFe] gives age and [M/H], \\
 \indent $\langle$Fe$\rangle$ vs [MgFe] gives [$\alpha$/Fe], and then\\
 \indent TiO$_{2{\rm SDSS}}$ vs [MgFe] gives the IMF.  \\
The panels on the right show that, while there are important differences between the MILES and C18 models, the same overall logic should still work for the C18 models, though not quite as cleanly.  Therefore, we now study how these SSP parameters, and the $M_*/L_r$ profiles they imply, differ for MILES and C18.

\subsection{SSP parameters and profiles}
To estimate age, [M/H], and [$\alpha$/Fe] profiles,  we fit the measured 
(emission corrected) H$_\beta$, $\langle$Fe$\rangle$, TiO$_{\rm 2SDSS}$ and [MgFe] line strengths to SSP models which span
 1–14 Gyr for age, 
 $-0.3$ to $0.2$~dex for [M/H], 
 $-0.4$ to $0.4$ for [$\alpha$/Fe] and 
 1.3–3.5 for the IMF slope parameters $\Gamma_b$ and $\alpha_1$. 
These SSP parameters can then be turned into profiles of $M_*/L_r$.

\begin{figure}
    \centering
      \includegraphics[width=0.85\linewidth]{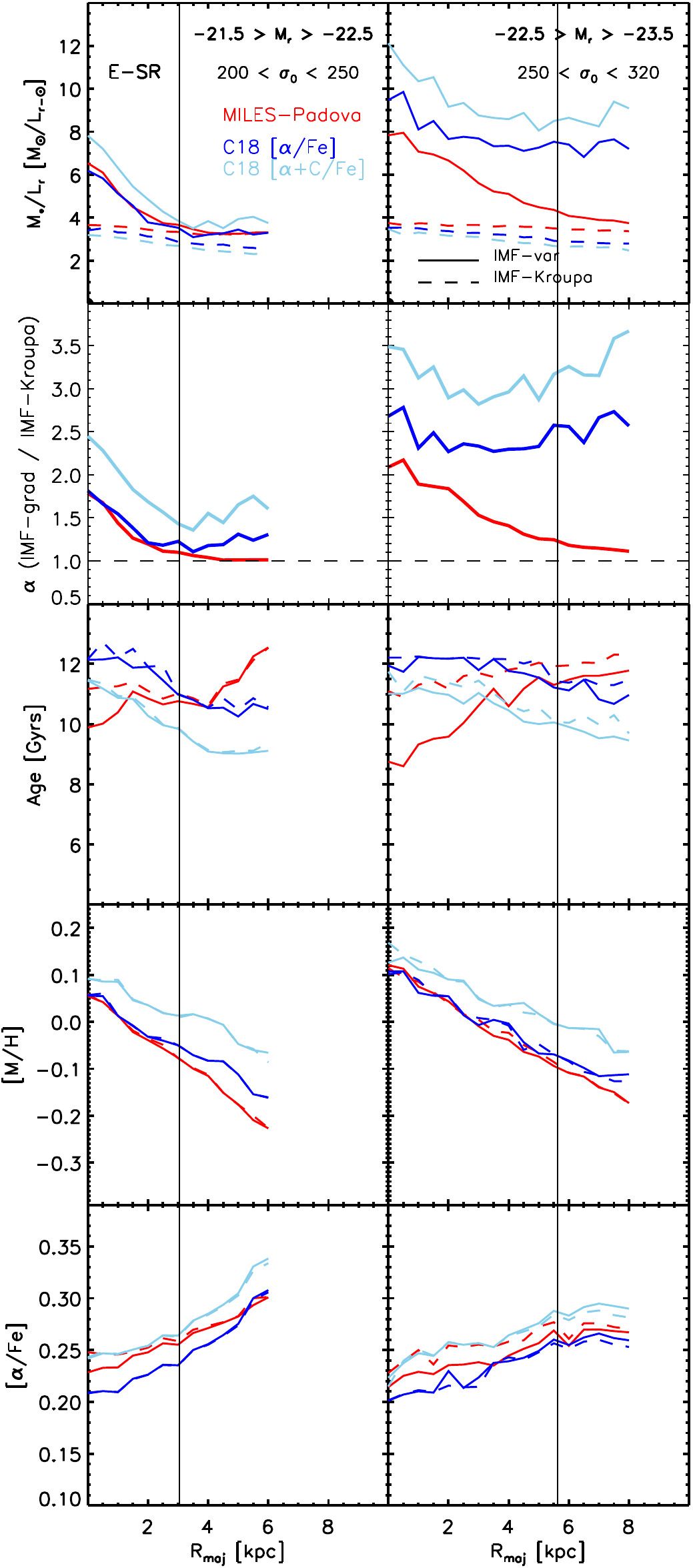} 
    \caption{Dependence of SSP parameters on the model (MILES+Padova, C18 or C18 with [C/Fe] enhancement) that is fit to the measured line-indices shown in Figure~\ref{fig:compare} for E-SRs in two bins in luminosity and velocity dispersion as labeled.   Solid curves result when the IMF is determined from the measurements along with the other parameters; dashed lines result if the IMF is fixed to Kroupa. }
    \label{fig:compareProfiles}
\end{figure}

Figure~\ref{fig:compareProfiles} shows the results of such an analysis.  Strictly speaking, the upper and lower limits on the H$_\beta$ line strength (see Section~\ref{sec:emierr}), which produce the hashed regions in the top panels of Figure~\ref{fig:compare}, translate to upper and lower limits on the various SSP parameters.  To avoid clutter, we show the SSP values which lie midway between these limiting values. (I.e., we do {\em not} fit the model grids using the midpoint of the hashed regions shown in Figure~\ref{fig:compare}.)  See Figure~\ref{fig:errors} in Section~\ref{sec:syserr} below for an idea of the actual size of the error bands.

The two sets of panels in Figure~\ref{fig:compareProfiles} show the two E-SR bins we studied in the previous figure (the yellow and red symbols).  For each, we show age, [M/H], and [$\alpha$/Fe] profiles, as well as profiles of $M_*/L_r$ and of the ratio of $M_*/L_r$ for the best fit IMF to that when the IMF is fixed to Kroupa (a ratio that, sadly, is also called $\alpha$, and about which we will have more to say in Section~\ref{sec:colors}).  In each panel, the different curves show profiles obtained from the MILES+Padova (red) and C18 models (blue), when we fix the IMF to Kroupa (dashed) and when the IMF is free to vary (solid).  In addition, because the C18 models allow one to enhance elements individually, we have included an analysis in which [C/Fe] can be enhanced in addition to [$\alpha$/Fe] (cyan). (We chose [C/Fe] in part because \citealt{Choi2019} believe that enhancing [C/Fe] is necessary to explain the broadband $g-r$ colors, a point to which we return in Section~4.3.)  For this case, we set [C/Fe]=[$\alpha$/Fe] and use equation~(\ref{eq:MH}) to determine the associated [M/H]. Appendix~A2 discusses enhancing [Ti/Fe] over and above its appearance in [$\alpha$/Fe].  

Except when [C/H] is enhanced, the two models produce very similar [M/H] profiles which decrease outwards and do not depend on the IMF.  They also have similar [$\alpha$/Fe] profiles (although C18 is slightly smaller) which tend to increase outwards, with little dependence on the IMF.  Thus, both models suggest that star formation occurred over a longer timescale in the central regions than at larger $R$ (although recall that these are light-weighted conclusions that are dominated by the younger stars).  Both MILES and C18 show that [M/H] and [$\alpha$/Fe] are anti-correlated.\footnote{It is worth noting that the spectra of individual stars in the solar neighbourhood show a clear anti-correlation between [M/H] and [$\alpha$/Fe] \cite[e.g.][]{Palla2022}.} 

The differences between MILES and C18 are larger for the quantities which more directly affect $M_*/L_r$:  ages and IMFs. The age profiles returned by the MILES analysis depend on the IMF:  if fixed to Kroupa throughout, then the ages are slightly younger in the inner regions, but this age gradient is slightly more dramatic when the IMF is allowed to vary. Larger age gradients when the IMF also varies are consistent with the discussion in the previous subsection, where we noted that bottom-heavy IMFs were required to explain the TiO$_{\rm 2SDSS}$ line strengths in the central regions. The younger ages in the central regions are qualitatively consistent with the more extended star formation histories suggested by the corresponding [$\alpha$/Fe] profiles for MILES.  They are also consistent with more emission in the central region if the emission is related to more recent star formation and not to gas lost during the later phases of stellar evolution \cite[recent near-ultraviolet spectroscopic studies, e.g.][have suggested that passively evolving massive, early-type galaxies host sub-one percent fractions of young stars in their innermost regions]{Salvador20,Salvador22}. In contrast, the C18 models have slightly older central regions whether the IMF is fixed to Kroupa or not. This may be in tension with the associated [$\alpha$/Fe] profiles; it is not obvious that one can have ages of order 12 Gyrs, and still accommodate extended star formation.

These stellar population parameters combine to predict the $M_*/L_r$ ratio.  (Strictly speaking, the tables of $M_*/L_r$ values provided by MILES do not include the effects of non-zero [$\alpha$/Fe], so we have computed them ourselves. The effects are small; see Appendix~B.)  
All the models show qualitatively similar gradients: $M_*/L$ decreases from the center outwards.  This decrease is small (less than 50\%) when the IMF is fixed to Kroupa, but much more dramatic (factors of 2 or more) when the IMF is allowed to vary.  This variation is generally confined to scales smaller than $R_e$; at larger scales the MILES models converge to the Kroupa value, but the C18 models may (top left panel) or may not (top right).  When $M_*/L_r$ does converge to Kroupa on large scales, the central value tends to be about a factor of two times larger.  Factor of two variations in $M_*/L_r$ have significant implications for studies of galaxy structure and dynamics which we explore elsewhere \citep{Bernardi2020, Bernardi2022}.  When the C18 profiles do not converge to Kroupa, they tend to be flat, and about two and a half to three times the Kroupa value.  If they do not drop to Kroupa, then the implied stellar mass would be two and a half to three times the Kroupa value, and would run the risk of being larger than the dynamical mass, which is unreasonable.  For this reason, we are more inclined to believe the C18 [$\alpha$/Fe] models than the ones in which [C/Fe] is also enhanced, and, if pressed to choose between C18 and MILES+Padova, we would choose the latter.

In summary, the MILES+Padova models produce reasonable ages; in contrast, the MILES+BaSTi models return ages that are unacceptably large.  
In addition, compared to MILES, the C18 models tend to return larger $M_*/L_r$ values, but sometimes return significantly smaller $M_*/L_r$ gradients.  When C18 gradients are small, their large $M_*/L_r$ ratios even at large $R$, potentially imply stellar masses which are unacceptably large.  The SSP parameters returned by MILES+Padova are the only ones which are not problematic over the entire range of galaxy types we have studied, so they are our fiducial choice.

\subsection{Systematic errors}\label{sec:syserr}
\begin{figure}
    \centering
      \includegraphics[width=0.85\linewidth]{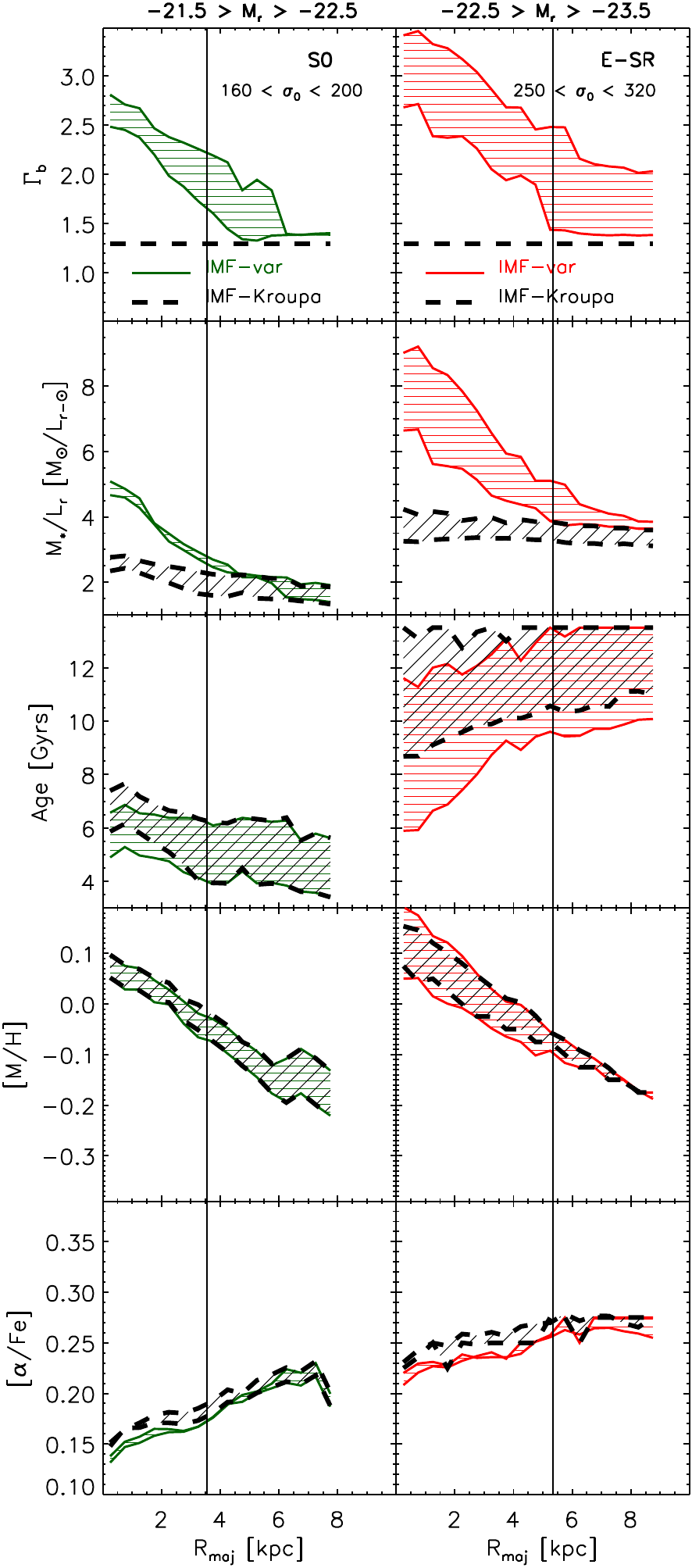} 
    \caption{Uncertainty on derived SSP parameters which result from the systematic uncertainty on the H$_\beta$ line strength shown in Figure~\ref{fig:compare}, when one uses the MILES+Padova models with IMF fixed to Kroupa (dashed) or allowed to vary (solid). Vertical line shows $R_e/2$.  Systematic errors are smaller for S0s (left) than E-SRs (right):  the much larger $M_*/L_r$ gradient when the IMF is allowed to vary than when it is fixed is robust to this systematic.}
    \label{fig:errors}
\end{figure}

We noted earlier that the statistical errors on our line-index measurements are negligible.  Therefore, uncertainties on our stellar population parameters are dominated by systematics.  If the underlying SSP model which we use to interpret the measurements is fixed (e.g. MILES+Padova), then these arise from, e.g., how we correct the measured spectra for emission.  To illustrate, the shaded regions in Figure~\ref{fig:errors} show the uncertainties in the MILES SSP parameter determinations which result from the shaded regions shown in Figure~\ref{fig:compare}:  the panel on the left is for the S0 bin (green), and the panel on the right is for the more luminous of the two E-SR bins (red).  The black shaded regions here result from fixing the IMF to Kroupa when estimating SSP parameters, and the colored regions are when the IMF is allowed to vary.  

\begin{figure}
    \centering
      \includegraphics[width=0.9\linewidth]{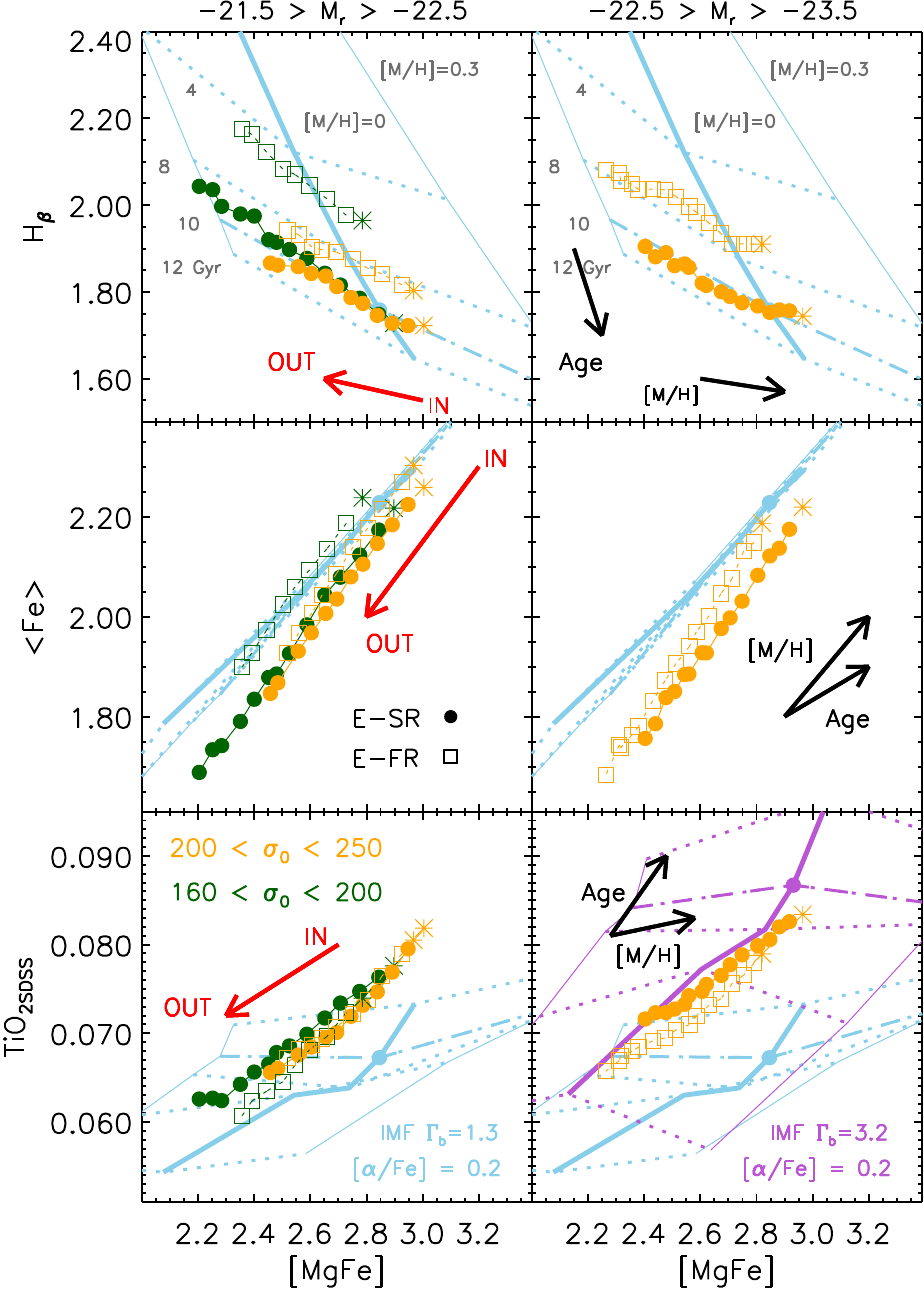} 
    \caption{Comparison of measured line index strengths of E-FRs (squares) and E-SRs (circles) in two luminosity (left and right) and $\sigma_0$ (yellow and green) bins.  Asterisks show the measurements in the central regions.  Each panel shows an age-metallicity SSP model grid from MILES+Padova when [$\alpha$/Fe]=0.2 and the IMF is Kroupa ($\Gamma_b\approx 1.3)$; bottom right panel shows one additional grid, for [$\alpha$/Fe]=0.2 and $\Gamma_b=3.2$.  Statistical errors are negligible; systematic errors (from the emission correction to H$_\beta$) are comparable to those shown in Figure~\ref{fig:compare}.} 
    \label{fig:types}
\end{figure}

Evidently, the uncertainty in the H$_\beta$ line strength drives uncertainties in the IMF, age and hence $M_*/L_r$ determinations (three upper panels), but much less so in [M/H] and [$\alpha$/Fe] (bottom two panels).  The widths of the uncertainty bands are largest for the most luminous E-SRs with the largest $\sigma_0$ since these are the ones for which the emission is hardest to detect, so the systematic error is largest.  Even in this case, however, the gradients in $M_*/L_r$ are obviously steeper when the IMF is allowed to vary than when it is fixed to Kroupa.  In addition, notice that in the left hand panel the uncertainty on $M_*/L_r$ is {\em smaller} when the IMF is allowed to vary than when it is fixed.  This indicates that the uncertainties on age and IMF -- which drive the uncertainties on $M_*/L_r$ -- are anti-correlated:  the upper limit on the ages (for which $M_*/L_r$ is largest) correspond to the lower limit on $\Gamma_b$ (for which $M_*/L_r$ is smallest).  

In Figure~\ref{fig:compareProfiles} and all the figures to follow (except for Figure~\ref{fig:types}), to avoid clutter we show the values which lie midway between the upper and lower limiting values at each $R$.  (In Figure~\ref{fig:types}, we show the values which correspond to the upper limit on H$_\beta$.)  Note that because the predicted line index strengths are nonlinear functions of the SSP parameters, fitting SSPs to the average of the two (differently corrected) line strength estimates returns SSP parameters which do {\em not} lie midway between the hashed regions shown.  

\section{Stellar population and $M_*/L$ gradients from MILES}\label{sec:miles}
We now turn to a study of how SSP parameters vary across the early-type galaxy population, subdivided by morphological type, when the IMF is allowed to be different for each stacked spectrum.  For reasons outlined at the end of the previous section, we only show results for the MILES+Padova SSP analysis.

\subsection{Line-index diagnostics and morphological type}
We begin with a comparison of the measured line indices and how they depend on luminosity (panels on right of Figure~\ref{fig:types} are bright), velocity dispersion (colors as labeled) and morphological type (symbols as labeled).  Cyan age-[M/H] grids in each panel are for [$\alpha$/Fe]=0.2 and a Kroupa IMF; purple grid in bottom right panel is for an enhanced model with bottom heavy IMF ([$\alpha$/Fe]=0.2 and $\Gamma_b=3.2$).

\begin{figure*}
    \centering
      \includegraphics[width=0.85\linewidth]{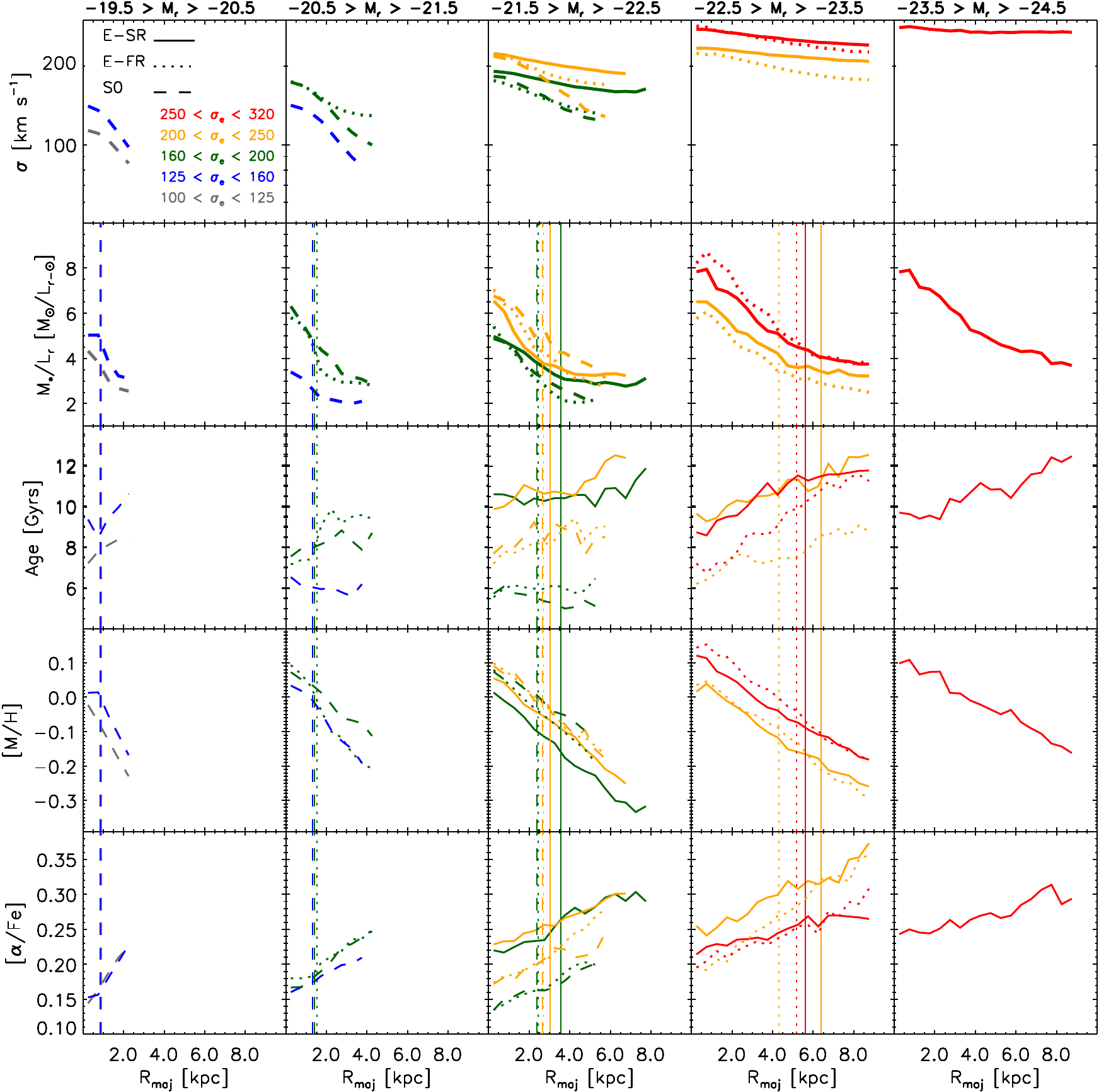} 
    \caption{Profiles of velocity dispersion and SSP-derived parameters $M_*/L_r$, age, total metallicity [M/H] and [$\alpha$/Fe] (top to bottom), when the line index combinations shown in Figure~\ref{fig:compare} are interpreted using the MILES+Padova models with the IMF left free to vary with scale.  Different columns show results for stacks constructed from objects with the specified absolute magnitude; colors indicate the value of $\sigma_0$; and line styles indicate morphology (solid, dotted, and dashed for E-SR, E-FR, and S0).  Vertical lines show $R_e/2$ of the objects in each stack.  Statistical errors are negligible; systematic errors are comparable to those shown in Figure~\ref{fig:errors}.  }
    \label{fig:miles}
\end{figure*}

Consider the right hand panels, which show that E-SRs (filled circles) have weaker H$_\beta$ (top) and slightly stronger [MgFe] (middle) and TiO$_{\rm 2SDSS}$ (bottom) than E-FRs (open squares) of the same $L$ and $\sigma_0$.  Comparison with the model grids in each panel suggests that the E-SRs are older (top), more $\alpha$-enhanced (middle) and, to match the TiO$_{\rm 2SDSS}$ line strength, the IMF must be more bottom-heavy than Kroupa.  Now consider the central spaxels, shown as asterisks.  These show that the TiO$_{\rm 2SDSS}$ line strengths of the E-SR and E-FR centers are very similar.  To be this similar, despite having different ages (as the top right panel suggests), requires that the E-FR IMFs must be even further from Kroupa than for E-SRs.  These extreme bottom heavy IMFs for E-FRs will have large $M_*/L$ values, which are balanced by the fact that they are younger (which tends to decrease $M_*/L$).  All these trends are also true in the panels on the left, and provide a simple way to understand the SSP profiles we present in the next subsection.  There we also comment on whether or not younger ages, small enhancements and extreme IMFs are physically reasonable.

\subsection{SSP profiles}
The bottom half of Figure~\ref{fig:miles} shows age, [M/H] and [$\alpha$/Fe] profiles determined by allowing the IMF to vary, as described in the previous section (i.e. by performing the same sort of MILES+Padova analysis as shown in Figures~\ref{fig:compare} and~\ref{fig:compareProfiles}), for a number of bins in $L$ (as labeled) and $\sigma_0$ (different colors) for S0s (dotted), E-FRs (dashed) and E-SRs (solid).  These, along with the best fit IMF, were used to estimate $M_*/L_r$ profiles which are shown in the panels which are second from top.  The top panels show the velocity dispersion profiles $\sigma(R)$ for these bins.   

Notice that $\sigma(R)$ depends on morphological type:  the profiles of E-FRs drop more steeply than those of E-SRs, and those of S0s drop even more steeply \cite[this is discussed in more detail by][]{DS2020}.  This is despite the fact that the shape of the profile was not used when making the morphological classification.  Before we discuss gradients in the stellar populations, we first highlight a few properties about the overall amplitudes.

The ordering of the line styles in the other panels shows that, at fixed $L$ and $\sigma_0$, E-SRs are older, slightly less metal rich and more $\alpha$-enhanced, compared to E-FRs and S0s and E-FRs.
In addition, at fixed $\sigma_0$, S0s and E-FRs with larger $L$ are younger, but this is not so obvious for E-SRs.  This is consistent with \cite{DS2019}.
At fixed $L$, the objects with larger $\sigma_0$ tend to have larger $M_*/L_r$.
Moreover, notice that in our next-to-highest luminosity bin (second from right hand panels), the two E-SRs (solid) have different $M_*/L_r$ even though they have similar ages.  Large $M_*/L_r$ is typically associated with older ages (as the aging population fades), whereas here $M_*/L_r$ appears to track [M/H] rather than age.  We show shortly that [M/H] and IMF shape are strongly correlated; this is consistent with our discussion in the previous subsection about $M_*/L$ being a competition between the age pushing it down (because H$_\beta$ is strong) and the IMF pushing it up (because TiO$_{\rm 2SDSS}$ is strong).  This (similar ages but different $M_*/L$) is also true for the central regions of the E-FRs (dotted).  

We now discuss gradients.  All objects have [M/H] decreasing outwards, in agreement with essentially all recent IFU-based analyses in e.g., SAURON, ATLAS3d, CALIFA, SAMI and MaNGA \cite[e.g.][]{Kuntschner2010, McDermid2015, califa2015, Santucci2020, Ge2021}.  
The age profiles are more complicated:  in most of our $L$ and $\sigma_0$ bins, E-SRs and E-FRs tend to be younger in the central regions.  While this is consistent with the tendency to have more emission in the center (c.f. Figure~\ref{fig:HbHaESR}), and is consistent with older work in the SAURON \citep{Kuntschner2010}, ATLAS3d \cite[][though the age gradient was not reported explicitly]{McDermid2015} and SAMI central galaxy \citep{Santucci2020} samples, it runs counter to a number of other studies suggesting that age decreases outwards: \citep{califa2015, Greene2015, Li2018, Parikh2019, Ge2021} in the CALIFA, MASSIVE and MaNGA surveys.\footnote{Although  \citealt{DS2019} did allow for IMF gradients and morphology, and three of the four curves in their Fig.16 show older ages in the central regions, we know that they depend on the emission correction; the bin for which the emission correction is smallest actually shows age decreasing towards the central regions.}  Note, however, that few of the studies which show age decreases outwards, and none of those which find it increases, allow for IMF gradients when determining SSP parameters.  In addition, none of these other studies could be as careful as us about morphology.   

As is the case for age profiles, the literature has yet to converge on the scale dependence of [$\alpha$/Fe].  All our [$\alpha$/Fe] profiles increase outwards (see also e.g., \citealt{Vaughan18}).  We believe that this, with the younger ages we see in the central regions, indicates more prolonged star formation there.  The age, [M/H] and [$\alpha$/Fe] gradients we see are broadly consistent with a `two-phase' formation scenario for these objects \cite[see, e.g., Section 5.4 of][and references therein, for details]{Santucci2020}.

The [$\alpha$/Fe] gradients are slightly weaker at large $\sigma_0$ (e.g. compare red and yellow curves), but it is not obvious that this extends to smaller $\sigma_0$.  Likewise, at large $L$ (bottom right panels) the amplitude of [$\alpha$/Fe] decreases as $\sigma_0$ increases, but this is not true for fainter E-SRs (bottom middle panel).  More work is needed to see if the two-phase process can explain these more detailed, morphology-dependent trends.

The associated $M_*/L_r$ profiles are peaked in the center, then drop by about a factor of two by $R=R_e/2$ before flattening beyond $R_e/2$.  (I.e., despite the differences in age gradients, the $M_*/L$ gradients are qualitatively similar to those in \citealt{DS2019}.)  Although the figure does not show this here (see Figure~8 in \citealt{Bernardi2022}), the large scale value is the same as when the IMF is restricted to Kroupa. 

Finally, it is worth highlighting the fact that some of the S0s and E-FRs are sufficiently young (e.g. $\le 7$~Gyrs) that stellar population estimates based on an SSP may not be a good approximation, especially given that they tend to have smaller [$\alpha$/Fe] values (which indicate star formation continued over a more extended period of time).   Thus, it is possible that the decision to describe these objects as SSPs leads to a systematic bias in the ages and, in particular, to the IMF.  In addition, if major mergers are important (as is widely assumed for E-SRs, e.g. \citealt{C16,Smith2020}), then they will tend to scramble stellar populations, so, unless the objects which merged had similar SSPs, treating them as SSPs may be unwarranted.  Since, ultimately, we are interested in $M_*/L$ gradients, we must worry if the SSP analysis biases us to overly large gradients.

To see that it does not, consider what would happen if we included a second component.  
If this component were younger and the primary component were older (than the SSP estimate), then matching the measured TiO$_{\rm 2SDSS}$ would still require a bottom heavy IMF (a young population with Kroupa IMF has small TiO$_{\rm 2SDSS}$).  Thus, a strong IMF gradient would still be required, and so, while the implied $M_*/L_r$ values would be modified slightly, the conclusion that a strong $M_*/L_r$ gradient is required would remain.

\subsection{Dependence on waveband}\label{sec:colors}
So far we have concentrated on $M_*/L_r$ gradients in the $r$-band.  However, as we discuss in detail in Appendix~B, the models also predict colors and hence $M_*/L_g$.  We discuss the colors here, and $M_*/L_g$ in \cite{Bernardi2022}.  

\begin{figure}
\centering
    \includegraphics[width=0.9\linewidth]{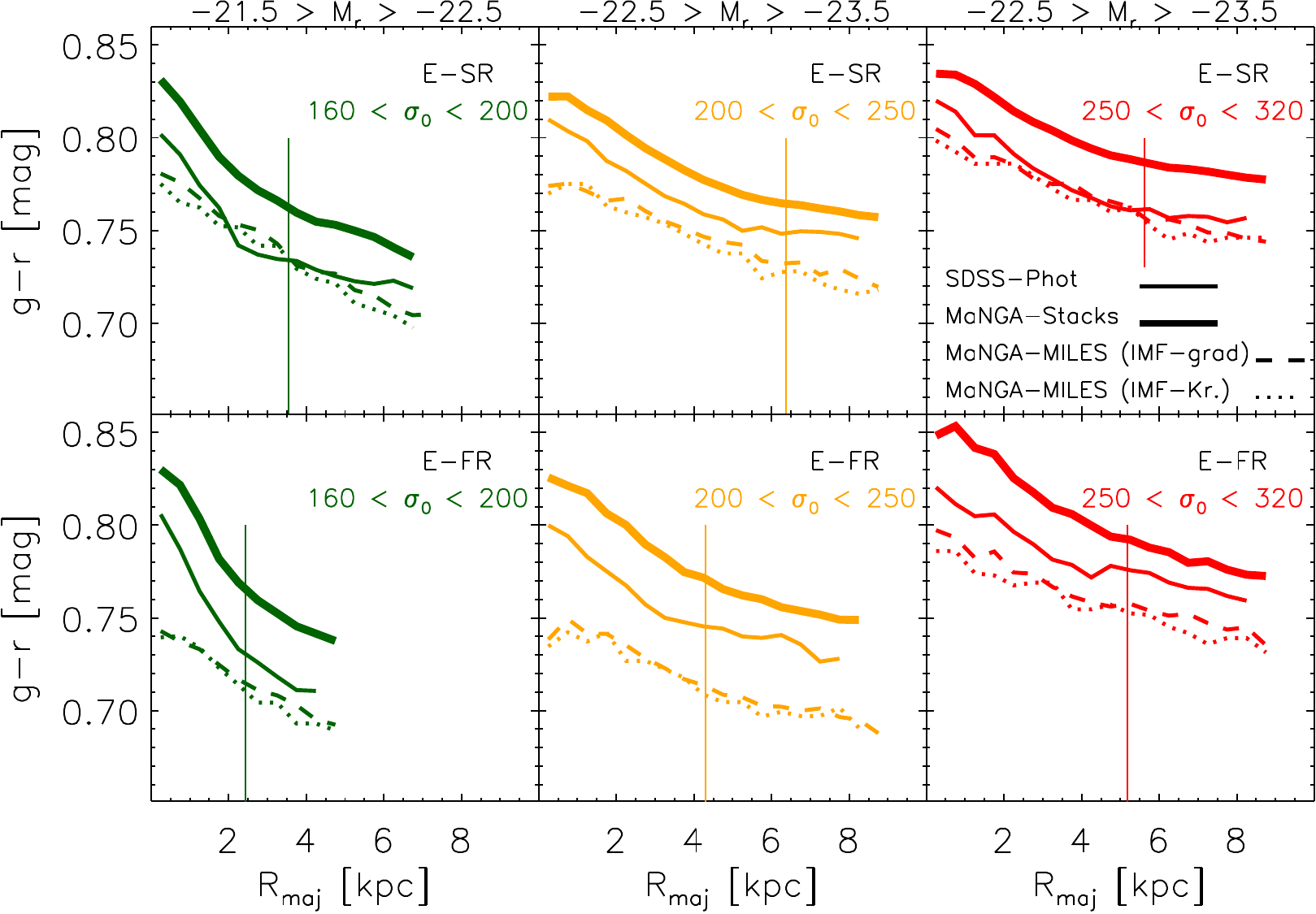}
    \caption{Comparison of restframe $g-r$ color profiles measured from the photometry (thin solid, $k$-corrected as described in the text) and from the stacked spectra (thick solid) for E-SRs (top) and E-FRs (bottom) for the $L$ and $\sigma_0$ values shown.  The offset indicates problems (e.g. flux calibration) which could complicate methods which use the full shape of the spectrum to constrain SSPs.  Dashed and dotted curves show the colors predicted by the best-fitting MILES+Padova SSP parameters to the H$_\beta$, $\langle$Fe$\rangle$, [MgFe], and TiO$_{\rm 2SDSS}$ line strengths.  In top panels, agreement with the thin solid curves indicates that SSPs determined from a few absorption lines are sufficient to predict the broad-band colors of E-SRs quite well. In bottom panels, models predict bluer colors than observed; stellar population analyses which assume an SSP may not be a good approximation for E-FRs (see text for details). }
    \label{fig:gmr}
\end{figure}

To compare the colors predicted by our MILES+Padova SSPs with those observed requires a few technical steps.  First, the spectroscopy.  The thick solid curves in Figure~\ref{fig:gmr} are constructed similarly to the `MaNGA-Stacks' curves in Figure~\ref{fig:gmr1by1}, except that now we show the restframe colors (we do not shift back to observed frame).  As a result, the bold curves here differ, primarily in amplitude, from those shown in Figure~\ref{fig:gmr1by1}.

Now for the photometry.  For each galaxy, we start with the observed-frame color profile.  Since the spectra were shifted to restframe, we must now `$k$'-correct the photometry to restframe as well.  For each $R$, we use a $k$-correction that we estimated by
 using the SSP values appropriate for the galaxy's $L$ and $\sigma_0$ (and $R$) to identify a model spectrum;
 measuring the flux that would be observed in the $g$- and $r$-bands to get the restframe color;
 then shifting the model spectrum to the galaxy's redshift, and remeasuring the flux in the $g$- and $r$-bands to get the observed frame color.  The $k$-correction, which is just the difference between these two color estimates, depends on $L$, $\sigma_0$, $R$ and $z$.  
 (In practice, this is less model dependent than it seems:  while a $k$-correction is necessary, it makes little difference if we use the one from MILES or from C18.)  Finally, each $k$-corrected galaxy contributes to the $g-r$ profile of the `restframe photometric stack' as many times as it contributed spaxels to the spectroscopic stack.  The thin solid curves (SDSS-Phot) in each panel show the color profiles which result.  Notice that they generally lie about 0.03~mags below (i.e. blueward) of the profiles from the spectra, for all morphological types.  This offset is similar to that shown in Figure~\ref{fig:gmr1by1}, so it is {\em not} due to problems with the $k$-correction.  

The two other curves in each panel of Figure~\ref{fig:gmr} show the restframe $g-r$ colors predicted by our MILES+Padova SSPs for each bin when the IMF is fixed to Kroupa (dotted) and when it is allowed to vary (dashed).  These colors were obtained by treating the predicted spectra just like the observed ones.  Notice that, in all panels, the dotted and dashed curves are very similar:  there is essentially no information about the IMF in the $g-r$ color.  

For the E-SRs (top panels), the models are within about 0.01~mags of the colors obtained from the photometry (SDSS-Phot).  This is remarkably good agreement, and indicates that SSPs determined from a few absorption lines are sufficient to predict the broadband colors quite well.  In particular, whereas the agreement between the models and the measured absorption lines shown in the left hand panels of Figure~\ref{fig:predict} are really a measure of goodness of fit -- since those were the lines to which the model was fit -- the thin curves shown here are genuine predictions, since broad band information is deliberately normalized out of Lick indices, so the $g-r$ predictions used {\em no} broadband information.  

The agreement is not nearly as good for the E-FRs (bottom panels) and S0s (not shown):  the predicted colors are bluer, sometimes substantially, than the photometry.  Recently, \cite{Choi2019} note that enhancing [C/Fe] or [N/Fe] will tend to make the predicted colors slightly redder.  However, as equation~(\ref{eq:MH}) shows, increasing [C/Fe] will increase the total [M/H] which, as is well known, will redden $g-r$ (see Appendix~\ref{sec:bbcolors}).  If this explains the reddening they see, then it is unlikely to explain the discrepancies we see.  Rather, we believe this is further indication that the stellar populations in these objects are more complex, so assuming that they are SSPs leads to biases (e.g. younger estimated ages and bluer colors).

\subsection{Correlations with [M/H]}
\begin{figure}
    \centering
    \includegraphics[width=0.95\linewidth]{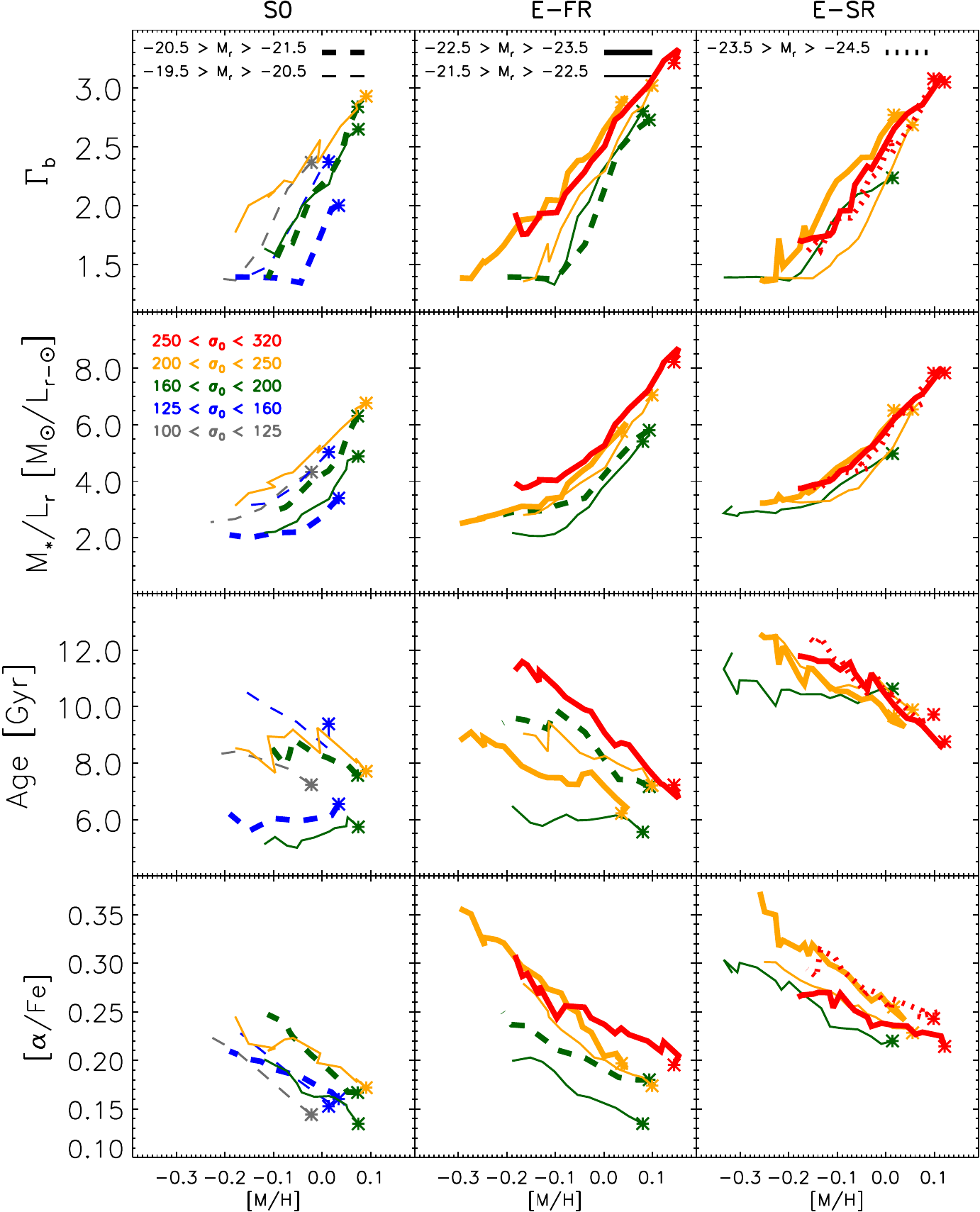}
    \caption{Correlations of the other SSP parameters with total metallicity [M/H], for S0s (left), E-FRs (middle) and E-SRs (right). Asterisks show the central point in each $L$ and $\sigma_0$ stack, and attached curves show larger $R$; line style indicates $L$ and color indicates $\sigma_0$.}
    \label{fig:MH}
\end{figure}

Figure~\ref{fig:MH} shows a different way of summarizing our results:  rather than showing gradients with respect to $R$, we show correlations with [M/H].  Since [M/H]-$R$ is typically a simple power law (c.f. Figure~\ref{fig:miles}), there is a sense in which the radial information is retained.  However, plotting versus [M/H] is interesting in view of the well known correlation between [$\alpha$/Fe] and [M/H] for individual stars in the Milky Way \cite[e.g.][]{Palla2022}, and many recent studies which argue that the IMF is correlated with \citep{MartinNavarro2015, Liang2021}, if not determined by \citep{Jevrabkova2018, Sharda2022} metallicity.

Each set of panels shows results for the three morphological classes:  S0s, E-FRs and E-SRs, and the line style and color of each curve in a panel indicate the $L$ and $\sigma_0$ bin of the stack from which the SSP parameters were derived.  In many respects, the results for E-SRs are simplest.  They show that IMF shape and $M_*/L_r$ are tightly correlated with [M/H], whereas age is anti-correlated (asterisks show the central region of each composite spectrum -- they have the highest metallicities); these trends are approximately the same for all $L_r$ and $\sigma_0$.  Presumably, the correlation of $M_*/L$ with [M/H] results because E-SRs are already quite old, so the effects of metallicity on $L_r$ dominate.  

The bottom panels show that, in all cases, [$\alpha$/Fe] and [M/H] are strongly anti-correlated, with the central regions of galaxies being least enhanced and most metal rich.  This is not unexpected: [$\alpha$/Fe] is a measure of the duration of star formation, so this says that star formation increases the total metallicity.  In our case, the zero-point of the anti-correlation depends on both $L$ and $\sigma_0$:  larger $L$ or larger $\sigma_0$ have larger [$\alpha$/Fe], but if one averages over $L$ at fixed $\sigma_0$, then [$\alpha$/Fe] increases with $\sigma_0$.  

For E-FRs, the qualitative trends are similar, but now there is more obvious dependence on $L$ and $\sigma_0$ even for $\Gamma_b$, $M_*/L$ and age.  
E.g., at fixed $\sigma_0$ the more luminous E-FRs are younger.  In addition, the E-FRs tend to be younger than E-SRs of the same $L$ and $\sigma_0$; objects with larger $\sigma_0$ tend to have larger [$\alpha$/Fe], but the dependence on $L$ for the two $\sigma_0$ bins is opposite.  This, along with the young ages, may be an indication that we are beginning to be biased by the assumption that the stellar population is an SSP; an analysis which allows for composite stellar populations may be required to obtain realistic estimates.

Panels which are second from the bottom show that, for E-FRs and E-SRs, age and [M/H] are anti-correlated:  the central regions are younger and more metal rich.  These trends are not as consistent for S0s.  Whereas E-SRs are old, and show little dependence on $L$ or $\sigma_0$, at fixed $\sigma_0$ the more luminous E-FRs are younger.  

It is not obvious that the tight correlations with [M/H] represent causal relationships.  In the simulations of \cite{Barber2019}, a correlation between IMF shape and [M/H] arises even though the IMF is not governed by metallicity at all.  While the `LoM' simulations of \cite{Barber2019} are in reasonable qualitative agreement with the gradients we see in MaNGA, they produce galaxies that are too young and insufficiently $\alpha$-enhanced.  We hope that our measurements motivate more such studies.

\section{Conclusions}
We studied the stellar populations of early-type galaxies in the MaNGA survey.  The optical colors of MaNGA early-type galaxies estimated from their spectra are slightly redder than the colors measured from the photometry (Figure~\ref{fig:gmr1by1}).  This redward tilt in the SED shape, which could be due to small flux calibration problems, potentially biases analyses which use the full SED shape to constrain SSP parameters.  In contrast, analyses which use Lick indices to constrain SSPs are much less sensitive to such broadband problems.  Lick index analyses require high S/N spectra which we achieve by stacking objects of similar morphology, $L$ and $\sigma_0$ (Figure~\ref{fig:SN}).  

Absorption line studies based on stacked spectra achieve substantially greater signal-to-noise if one normalizes each spectrum by the (index dependent) pseudo-continuum prior to stacking (Figure~\ref{fig:normHb}).  Our stacks of spectra around the H$_\beta$ line show that there is more emission in the central regions of E-SR and E-FR galaxies than at larger $R$ (Figures~\ref{fig:HbHaLine} and \ref{fig:HbHaESR}). This may indicate that the stars in the central regions are slightly younger, making it likely that they are also less $\alpha$-enhanced.  A comparison of the H$_\beta$, $\langle$Fe$\rangle$, TiO$_{\rm 2SDSS}$ and [MgFe] absorption line strengths -- Appendix~\ref{sec:which} discusses why we use these four lines rather than others -- with the predictions of the MILES SSP models suggests that this is indeed the case (Figures~\ref{fig:compare} and~\ref{fig:miles}). In contrast, total metallicity [M/H] decreases as one moves outwards from the central regions.  These absorption line strengths, and their gradients, correlate with morphology, which the MILES models interpret as indicating that E-SRs are older and formed their stars on a shorter timescale than E-FRs or S0s (Figure~\ref{fig:miles}).  The youngest E-FRs have slightly anomalous [$\alpha$/Fe] values (Figures~\ref{fig:miles} and~\ref{fig:MH}); analyses based on composite stellar populations may be more appropriate for such objects.

Although not the primary focus or our study, it is worth noting that the gradients we see, expecially for the E-SRs, are consistent with a `two-phase' formation process, in which an initial phase of rapid in situ star formation in what is now the inner regions (with more extended star formation towards the center, producing younger, more metal rich, and lower $\alpha$-enhanced stars in the center) is followed by a second phase in which mass is assembled via dissipationless mergers.  (See, e.g., Section 5.4 of \citealt{Santucci2020} for a more detailed discussion, but note that our measurements really only probe scales smaller than $R_e$.)  More work is needed to see if this scenario can explain the complex dependence of the amplitudes (overall age, [M/H] and [$\alpha$/Fe]) on $L$, $\sigma_0$ and morphology that we see in Figure~\ref{fig:miles}.

One reason we have not pursued this further is that, beacuse of systematic differences between the MILES and C18 models (Figure~\ref{fig:compare}), the same observed line strength gradients imply different age gradients in the C18 models (Figures~\ref{fig:compareProfiles} and~\ref{fig:response} and associated discussion).  Even for the simplest case of solar abundances and a Kroupa IMF, the model age-metallicity grids are different, especially for H$_\beta$-[MgFe], but also for TiO$_{\rm 2SDSS}$-[MgFe].  This leads to age-[M/H] differences which propagate to other SSP parameters.  

Moreover, the way in which these grids respond to variations in $\alpha$-element abundances also differs, especially for H$_\beta$-[MgFe].  There are also significant differences in the predicted response to changes in the IMF (Figures~\ref{fig:compareProfiles}).  It is tempting to conclude that this is driven by differences in how the IMF is parameterized (Figure~\ref{fig:imfs}): The MILES parameterization leads to a smaller gas fraction (Figure~\ref{fig:fgas}), hence a larger mass fraction in low mass stars and remnants when the IMF is bottom heavy (compared to Kroupa).  However, the models predict different responses even for similar IMFs (e.g. the Salpeter versus Kroupa grids in Figure~\ref{fig:compare}).  

Differences in inferred SSP parameters imply $M_*/L_r$ values that are also different:  if the IMF is held fixed to Kroupa then the MILES values are larger, but if the IMF is allowed to vary, the C18 values are larger (Figure~\ref{fig:compareProfiles}, although the IMF differences in this figure are due, in part, to differences in how the IMF is parameterized).  These differences impact the implied $M_*/L_r$ gradients.  Allowing bottom-heavy IMFs in MILES yields $M_*/L_r$ values that can exceed the Kroupa values by a factor of two in the central regions, but they become the same as the Kroupa values beyond about half the half light radius (Figure~\ref{fig:miles}).  For C18, the excess on small scales can be even larger, and in some cases the values at large $R$ can remain 2 or 3 times larger than Kroupa at scales of order $R_e$ (e.g. top right panel of Figure~\ref{fig:compareProfiles}).  If this persists to even larger $R$, then the estimated $M_*$ would exceed the dynamical mass, which is unrealistic.

Optical colors are extremely insensitive to the IMF (Figure~\ref{fig:gmrAlpha}); only when the IMF is extremely bottom-heavy (the IMF slope is large) is there a small effect:  in this limit, if the age and [M/H] are fixed, then increasing the IMF slope as one increases [$\alpha$/Fe] can keep the color unchanged.  Even though the absorption lines we interpret using SSPs have no broad-band information, the MILES+Padova SSP parameters inferred from these lines are able to predict the broad-band optical ($g-r$) colors to within about 0.01~mags (Figure~\ref{fig:gmr}).  

\cite{Bernardi2022} use the $M_*/L_r$ gradients determined from our MILES+Padova analysis to quantify the difference between the projected half light and half mass radii.  However, because the gradients associated with C18 are different (Figure~\ref{fig:compareProfiles}), $R_{e*}/R_e$ will also be different.  This is mainly driven by the systematic differences between the MILES and C18 age-metallicity grids shown in Figure~\ref{fig:compare}.  
The unprecedented signal-to-noise of our measurements (Figure~\ref{fig:SN}) means that these systematics in the models are now too large to be ignored.  We hope that our study motivates future work aimed at resolving these differences, which are present even for the simplest case of solar abundances and a Kroupa IMF.  This is especially necessary in this era of canned routines which allow one to analyze many lines at once, or fit the entire spectrum, but also make it too easy to forget that there are important unsettled systematic differences hiding under the lid. 

\section*{Acknowledgements}
We are grateful to K. Westfall for clarifications about changes in the MaNGA database between DR15 and DR17, to C. Conroy and A. Vazdekis for discussion of their SSP models, and to the anonymous referee for a helpful report which improved this paper. This work was supported in part by NSF grant AST-1816330.

Funding for the Sloan Digital Sky Survey IV has been provided by the Alfred P. Sloan Foundation, the U.S. Department of Energy Office of Science, and the Participating Institutions. SDSS acknowledges support and resources from the Center for High-Performance Computing at the University of Utah. The SDSS web site is www.sdss.org.

SDSS is managed by the Astrophysical Research Consortium for the Participating Institutions of the SDSS Collaboration including the Brazilian Participation Group, the Carnegie Institution for Science, Carnegie Mellon University, the Chilean Participation Group, the French Participation Group, Harvard-Smithsonian Center for Astrophysics, Instituto de Astrof{\'i}sica de Canarias, The Johns Hopkins University, Kavli Institute for the Physics and Mathematics of the Universe (IPMU) / University of Tokyo, Lawrence Berkeley National Laboratory, Leibniz Institut f{\"u}r Astrophysik Potsdam (AIP), Max-Planck-Institut f{\"u}r Astronomie (MPIA Heidelberg), Max-Planck-Institut f{\"u}r Astrophysik (MPA Garching), Max-Planck-Institut f{\"u}r Extraterrestrische Physik (MPE), National Astronomical Observatories of China, New Mexico State University, New York University, University of Notre Dame, Observat{\'o}rio Nacional / MCTI, The Ohio State University, Pennsylvania State University, Shanghai Astronomical Observatory, United Kingdom Participation Group, Universidad Nacional Aut{\'o}noma de M{\'e}xico, University of Arizona, University of Colorado Boulder, University of Oxford, University of Portsmouth, University of Utah, University of Virginia, University of Washington, University of Wisconsin, Vanderbilt University, and Yale University.

\section*{Data availability}

The data underlying this article are available in the Sloan Digital Sky Survey Database at https://www.sdss.org/dr17/.





\bibliographystyle{mnras}
\bibliography{biblio} 




\appendix

\section{Choice of absorption lines}\label{sec:which}

In the main text we noted that the H$_\beta$-[MgFe] combination is attractive because, in this plane, age and metallicity are nicely separated and almost independent of $\alpha$-element abundances (at least for the MILES models). 
As we now discuss, for the MILES models, other line-index combinations which provide constraints that are relatively straightforward to visualize are
$\langle {\rm Fe}\rangle$-[MgFe], which provides a nice diagnostic of $\alpha$-element abundances; 
and the TiO$_{\rm 2SDSS}$-[MgFe] combination, which constrains the IMF.

\begin{figure*}
  \centering
  \includegraphics[width=0.8\linewidth]{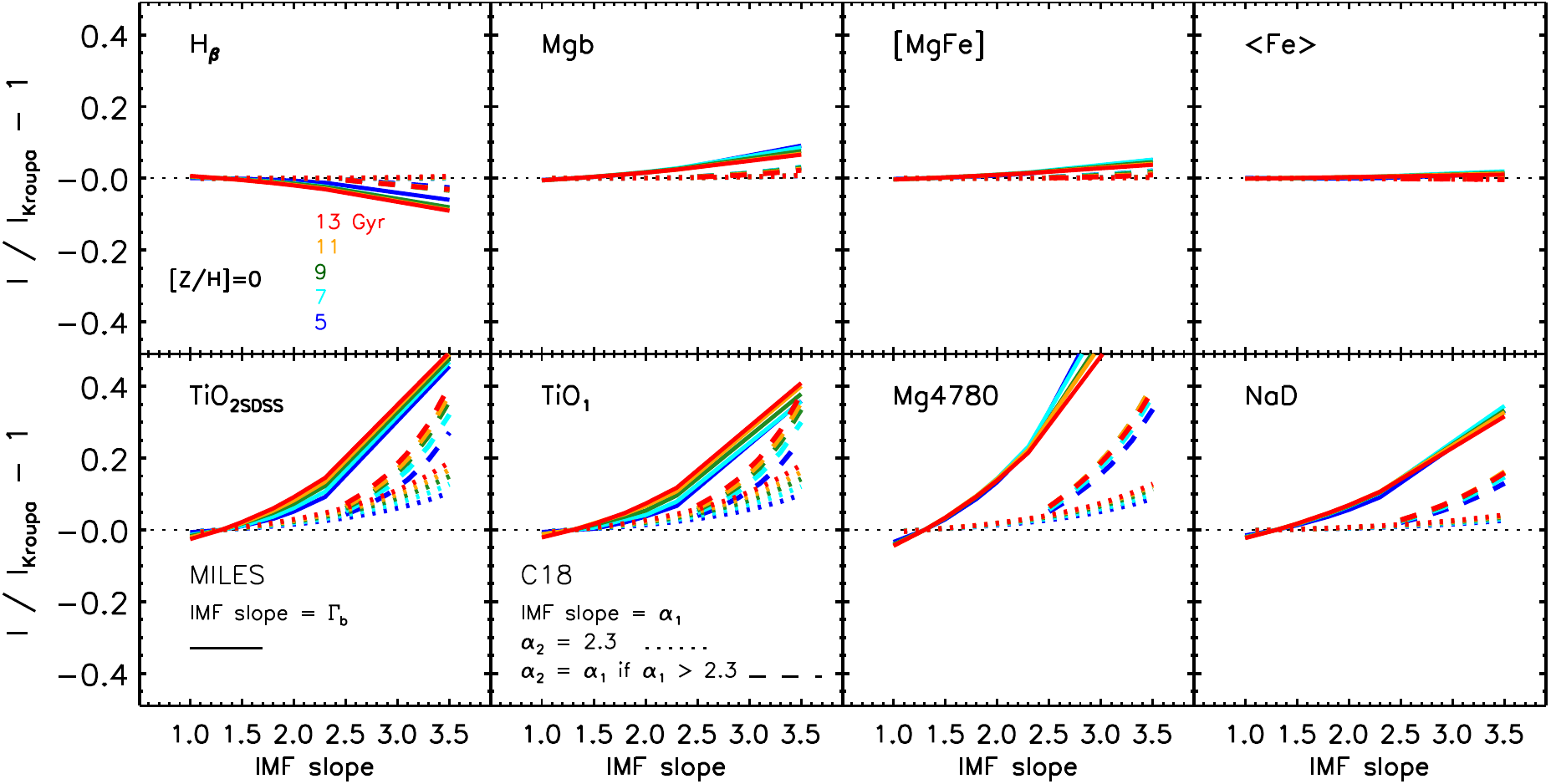}
  \caption{Response of various line index strengths to changes in the IMF (as labeled), for various ages at [Z/H]=0 (i.e. solar metallicity and abundance ratios). The indices in the bottom panel are more sensitive to the IMF.  }
  \label{fig:responseIMF}
\end{figure*}

\subsection{Element abundances and the IMF}
Since we are primarily interested in IMF-related effects, the top panels in Figure~\ref{fig:responseIMF} show lines that are commonly used to estimate age, metallicity and $\alpha$-enhancement, and the bottom panels show lines that are more sensitive to the IMF.  Each panel shows how line strength varies with IMF shape for a range of ages when [Z/H]=0, normalized by its strength when the IMF is Kroupa. Solid lines show the predicted strengths from the MILES+Padova SSPs, for which the `bimodal' IMF shape is determined by one free parameter, $\Gamma_b$.  The dashed and dotted lines show results from C18, which, in our implementation, also have one free parameter $\alpha_1$:  the other parameter, $\alpha_2$, is set to either 2.3 (dotted) or $\alpha_1$ when $\alpha_1> 2.3$ (dashed).  Despite the fact that the Kroupa IMF happens to have $\Gamma_b=1.3$ and $\alpha_1=1.3$, the IMF shapes for other choices are quite different, so we do not expect the line strengths when $\Gamma_b=2.3$, say, to be the same as when $\alpha_1=2.3$.  The main point of this plot is that the lines in the top panels vary weakly, if at all, with IMF, whereas the indices in the bottom are much more sensitive to IMF changes \citep{LaBarbera2013, Vazdekis2015, TW2017, LaBarbera2019, DS2019}.  

\subsection{Element abundances and enrichment}
We now study systematic differences between MILES and C18 when the IMF is fixed to Kroupa but element abundance ratios are allowed to vary.  We use this study to motivate which subset of the lines shown in Figure~\ref{fig:responseIMF} to use in the main text.  

In MILES, non-solar element abundance ratios are controlled by a single quantity, [$\alpha$/Fe], where $\alpha$ is a particular combination of O, Ne, Mg, Si, S, Ca and Ti.  For this quantity, equation~(\ref{eq:MH}) relates the total metallicity [M/H] to that when the abundances are solar, [Z/H] (recall that [C/Fe]=0 in MILES).  To gain intuition, we will be interested in varying some of these elements individually.  This is not an option in MILES, which only allows variations in [$\alpha$/Fe] as a whole, but the C18 models allow one to enhance some elements individually while keeping [Z/H] fixed.  

To implement enhancements, the C18 models provide `response functions' which describe how spectra of a given [Z/H] are modified by an enhancement of 0.3~dex (except for C, in which case 0.15~dex), where the enhancement is in [X/H]$_{\rm rf}$ for all elements X except for the quantity $\alpha_{\rm S}$, a combination of (O, Ne, S), which is provided as [$\alpha_{\rm S}$/Fe]$_{\rm rf}$ (i.e. an enhancement with respect to Fe rather than H).  The subscript `rf' is necessary because the library spectra themselves have non-solar abundances [X/Fe]$_{\rm lib}$ which depend on [Z/H] (provided in Table~1 of C18, and recall that their [Fe/H] is our [Z/H]), and must be accounted for.  Unfortunately, C18 do not provide values for [$\alpha_{\rm S}$/Fe]$_{\rm lib}$.  Of the elements which contribute to $\alpha_{\rm S}$, only O has an estimated [O/Fe]$_{\rm lib}$.  Although we could use the solar mass fractions of O, Ne and S to define
 ${\rm [\alpha_S/Fe]_{lib}} = \log_{10} (0.22 + 0.78\,10^{\rm [O/Fe]_{lib}})$,
we instead set 
  [$\alpha_{\rm S}$/Fe]$_{\rm lib}$ = [O/Fe]$_{\rm lib}$,
since it is very likely that Ne and S track O.  (Since O dominates, the difference is small.)

So, suppose we wish to build a spectrum which represents a population with a given [Z/H] and [$\alpha$/Fe] (and age and IMF) values, starting from C18 spectra with given [Z/H] and [$\alpha_{\rm S}$/Fe]$_{\rm rf}$ values.  Then we must determine the appropriate values of [X/H]$_{\rm rf}$ such that [X/Fe] = [$\alpha_{\rm S}$/Fe] for each of X = Mg, Si, Ca and Ti.  Since we want 
\begin{equation}
 {\rm [X/Fe]} 
  = {\rm [X/H]_{rf}} - {\rm [Fe/H]_{rf}} + {\rm [X/Fe]_{lib}}
  = {\rm [\alpha_S/Fe]} ,
\end{equation}
we have
\begin{equation}
 {\rm [X/H]_{rf}} = {\rm [\alpha_S/Fe]} - {\rm [X/Fe]_{lib}} + {\rm [Fe/H]_{rf}}.
\end{equation}

\begin{figure*}
  \centering
  \includegraphics[width=0.8\linewidth]{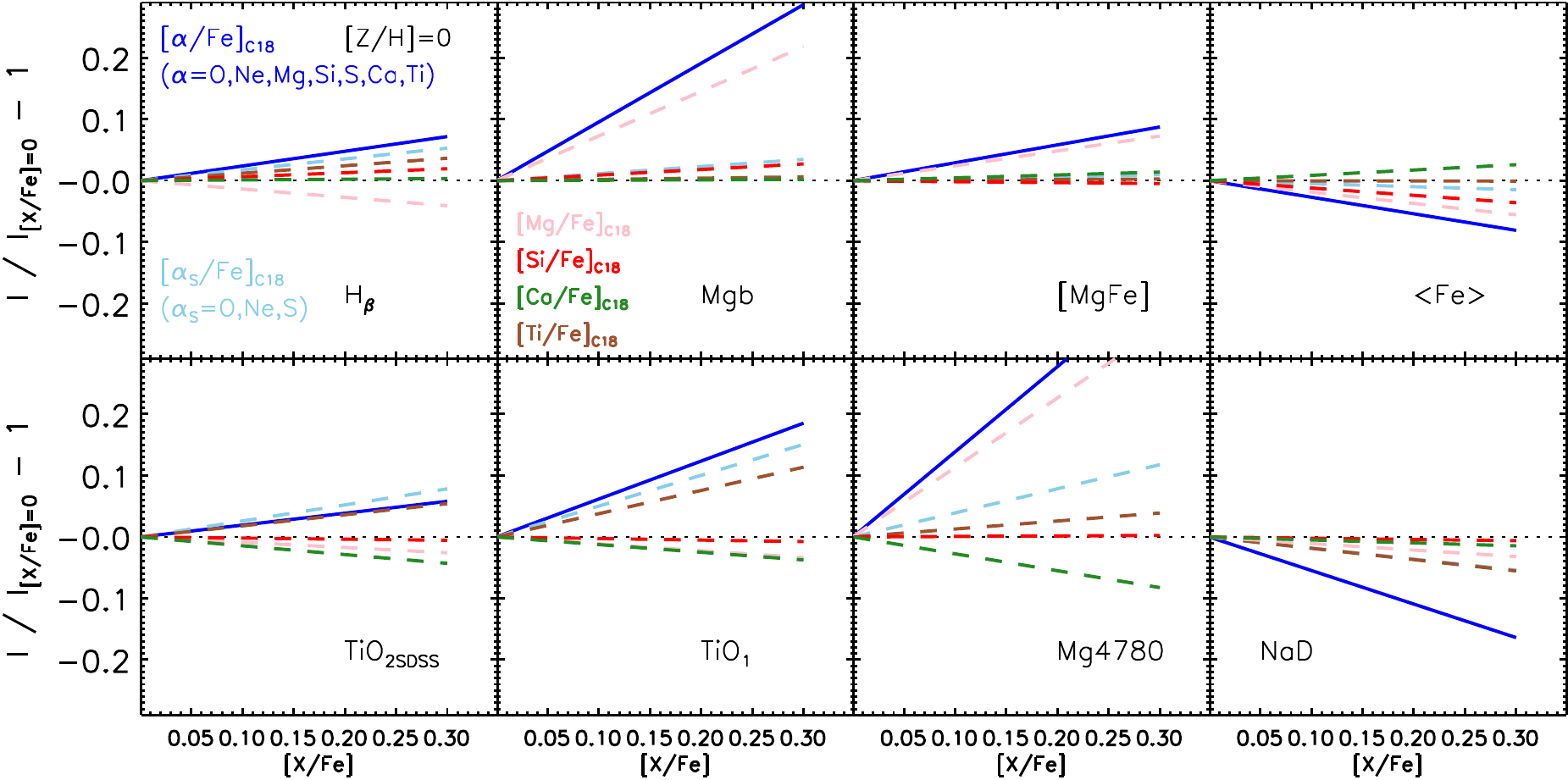}
  \caption{Response of various indices (as labeled) to changes in a few $\alpha$-element abundances relative to solar (different lines in each panel) for a 9~Gyr old SSP with [Z/H]=0 and a Kroupa IMF in the C18 models.}
  \label{fig:C18responses}
  \includegraphics[width=0.8\linewidth]{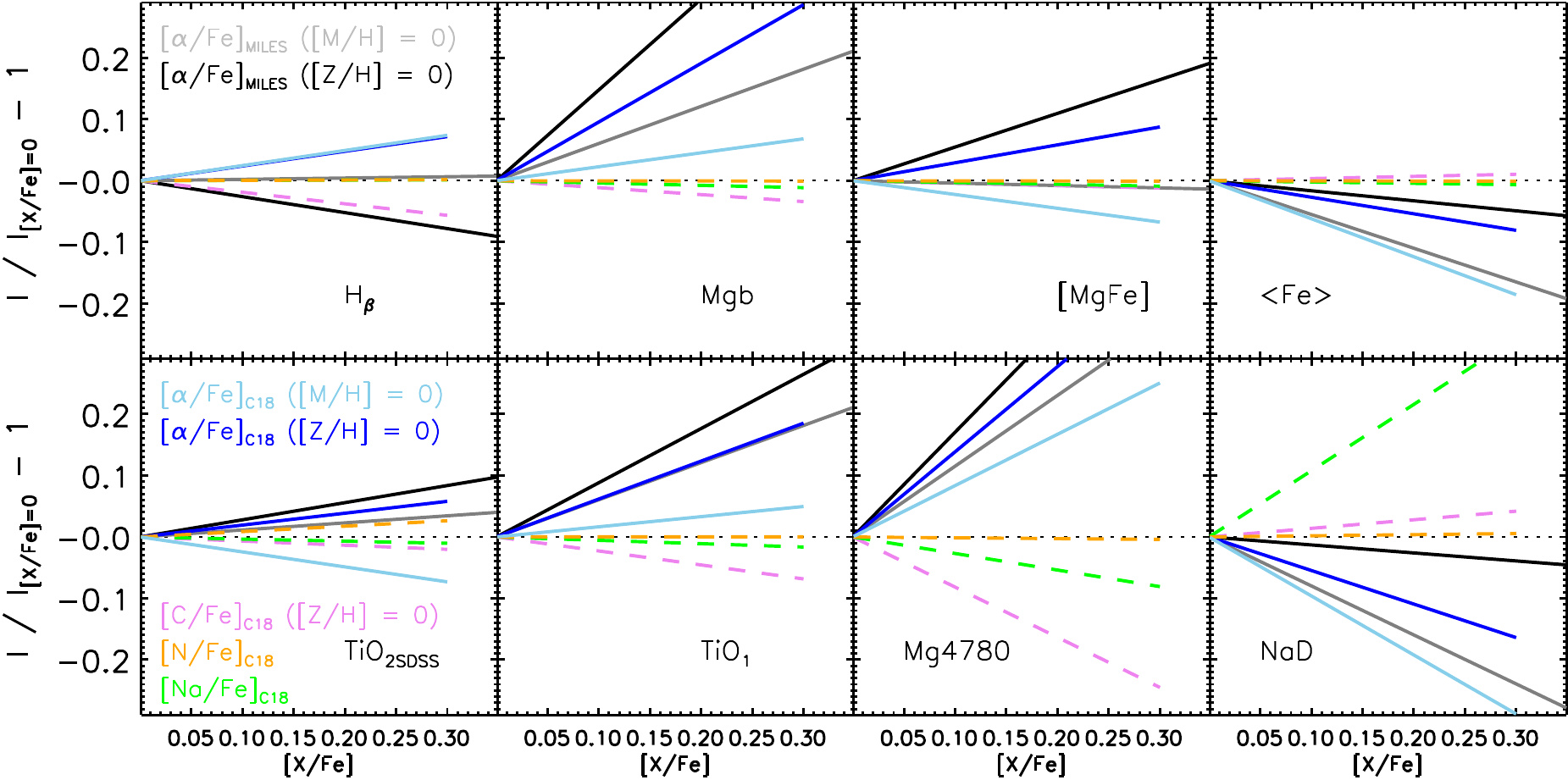}
  \caption{Same as previous figure, but now showing C18-predicted responses to enhancements in a few other elements, as well as responses predicted by MILES to variations in [$\alpha$/Fe].  Except for the two cases with [M/H]=0, all the other variations are with respect to [Z/H]=0.}
  \label{fig:response}
\end{figure*}

In practice, we assume that [Fe/H]$_{\rm rf}$=0.  I.e., we assume that C18 achieve [$\alpha_{\rm S}$/Fe]$_{\rm rf}$=0.3 by enhancing O, Ne and S, and not by depleting Fe (with respect to solar abundance ratios), despite the discussion in \cite{Trager2000} and~\cite{TMB2003}.

The corresponding spectrum is
\begin{equation}
  {\cal F}_{\rm non-solar}(\lambda) = {\cal F}^{\rm [Z/H]}_{\rm empirical}(\lambda) \prod_i R_i(\lambda)
\end{equation}
where the $i$ labels the elements as well as $\alpha_{\rm S}$ and 
\begin{equation}
 R_i = 1 \pm \left(\frac{{\cal F}^{\rm [Z/H]}_{{\rm X}_i}}{{\cal F}^{\rm [Z/H]}_{\rm Solar}} - 1\right)\, \frac{{\rm [X}_i{\rm /H]_{\rm rf}}}{0.3}.
\end{equation}
The plus sign is used if [X$_i$/H]$>0$, the minus sign otherwise,
the [Z/H] super-scripts indicate that $R_i$ is correcting a spectrum with the given value of [Z/H] (in addition to age and IMF) but solar abundance ratios, and the 0.3 in the denominator is because C18 only provide spectra for enhancements of $\pm 0.3$~dex.  (For $\alpha_{\rm S}$, [X$_i$/H]$_{\rm rf}$$\to$[$\alpha_{\rm S}$/Fe]$_{\rm rf}$.)
The total metallicity [M/H] is then given by
\begin{equation}
 {\rm [M/H]} = {\rm [Z/H]} + \log_{10} \left(Z_0 + \sum_i Z_i\, 10^{{\rm [X_i/Fe]}}\right),
\end{equation}
with $Z_0 = 1-\sum_i Z_i$ (see equation~\ref{eq:MH}).
To enhance [C/Fe] in addition to [$\alpha$/Fe] one would include a response term $R_{\rm C}$ with [C/H]$_{\rm rf}$ and $0.3\to 0.15$, and one would include the [C/Fe] contribution to [M/H].

The blue line in the top left panel of Figure~\ref{fig:C18responses} shows how H$_\beta$ responds as [$\alpha$/Fe] is increased while keeping [Z/H] fixed.  (We use the subscript C18 in all cases to remind the reader that, e.g., [$\alpha$/Fe]$_{\rm C18}$ is constructed using the C18 models.)  Although we only show results for [Z/H]=0, they are qualitatively similar for other values.  The cyan dashed line just below it shows how H$_\beta$ responds as the subset [$\alpha_{\rm S}$/Fe] comprising [O, Ne, S] is enhanced, and the other dashed lines in the panel show how H$_\beta$ responds as [Mg/Fe], [Si/Fe], [Ca/Fe] and [Ti/Fe] are enhanced individually.  Evidently, enhancing [Mg/Fe] decreases H$_\beta$, but this can be canceled if the other three are also similarly enhanced, leaving [$\alpha_{\rm S}$/Fe]$\approx$ [$\alpha$/Fe].

The other panels on the top row of Figure~\ref{fig:C18responses} show how the Mg$b$, $\langle$Fe$\rangle$ and [MgFe] lines respond to these changes.  Evidently, [MgFe] is almost completely determined by [Mg/Fe], and enhancing [$\alpha$/Fe] actually decreases $\langle$Fe$\rangle$.  The bottom panels of Figure~\ref{fig:C18responses} show how a number of other indices respond to these changes.  These include TiO$_{\rm 2SDSS}$ (which we use in the main text) and TiO$_1$, Mg4780 and NaD (which we do not, for reasons we discuss below).  Evidently, one can enhance [Si/Fe] without affecting any of the lines in the lower panels, while fine-tuning the lines in the upper panel.  Thus, varying elements individually might allow one to fit a given collection of line strengths better.

The C18 models also allow one to explore responses to enhancements in elements that are not included in the MILES definition of [$\alpha$/Fe]:  the dashed lines in Figure~\ref{fig:response} show how these indices vary as [C/Fe], [N/Fe] and [Na/Fe] are enhanced.  This shows, for example, that one can use [Na/Fe] to adjust the predicted NaD line strength without affecting any of the lines in the upper panels, or the two TiO lines.  Similarly, the dashed pink lines show how all these lines respond to [C/Fe].  Enhancing [C/Fe] reduces TiO$_1$ and Mg4780, so one can balance the increase in TiO$_1$ as [$\alpha$/Fe] increases by simultaneously increasing [C/Fe].  Increasing [C/Fe] will have a small effect on TiO$_{\rm 2SDSS}$ and Mg$b$, but almost none on [MgFe] and $\langle$Fe$\rangle$.  As a result, one can fix most of the discrepancy between the two Ti lines by enhancing [C/Fe], provided one can accommodate the change in the age estimate associated with the fact that H$_\beta$ responds to [C/Fe].

Likewise, one could explore the effects of further enhancing any of the elements that are included in the MILES definition of [$\alpha$/Fe].  The brown dashed lines show that varying [Ti/Fe] makes no difference to [MgFe] or $\langle$Fe$\rangle$, but can increase both TiO$_{\rm 2SDS}$ and H$_\beta$.

Having established how one can, in principle, enhance elements one by one, we now ask if this (time consuming exercise) is worth doing in practice.  The question is motivated by the fact that, if there are systematic differences between MILES and C18, then one must interpret the results of any such exercise with caution.  

We can only compare the C18 and MILES responses to non-solar abundances for [$\alpha$/Fe], and even then we must take care to account for the fact that MILES works with fixed metallicity [M/H] whereas C18 keeps [Fe/H] fixed.  Since we have been considering [Z/H]=0, equation~(\ref{eq:MH}) indicates that enhancing [$\alpha$/Fe] requires [M/H]$<0$.  The black lines in each panel show [$\alpha$/Fe] in MILES when [Z/H]=0; unfortunately, they are different from the blue lines which show the same enhancement but for C18.  The two predict very different responses for H$_\beta$, and are moderately different for the other indices we use in the main text.  This, fundamentally, is why the two models make different predictions for SSP parameters.  

To see this more directly, we now consider the responses when we keep [M/H] fixed -- this is the more natural way to present the MILES models.  We will show [M/H]=0, but our argument is general.  We again use equation~(\ref{eq:MH}), but this time to work out what value of [Z/H] to use for the C18 models as [$\alpha$/Fe] increases.  The grey line in each panel shows the index strengths predicted by MILES, and the cyan line shows C18.  Consider the grey lines first.  These show that H$_\beta$ and [MgFe] are unchanged as [$\alpha$/Fe] increases.  This is why H$_\beta$-[MgFe] is a good diagnostic of age and total metallicity [M/H] but $\langle$Fe$\rangle$-[MgFe] is a good measure of [$\alpha$/Fe] \citep{Thomas2005}.  Similarly, the grey curves in the bottom panels show that, in the MILES models, TiO$_{\rm 2SDSS}$ is rather insensitive to [$\alpha$/Fe], whereas the other lines are all quite sensitive to it.  Consequently, for the MILES models, these other lines are not as straightforward probes of IMF effects as TiO$_{\rm 2SDSS}$, so we use TiO$_{\rm 2SDSS}$-[MgFe] as a probe of the IMF in the main text.

Now consider the cyan curves.  These show that [MgFe] for C18 is weakly sensitive to [$\alpha$/Fe], whereas H$_\beta$ is quite sensitive, in stark contrast to MILES.  This means that MILES and C18 will require different ages and enhancements to explain the line strengths we measure in MaNGA.  In addition, compared to MILES, the TiO$_{\rm 2SDSS}$ line in C18 is more sensitive to [$\alpha$/Fe], but in the opposite sense.  Moreover, whereas in MILES, both TiO$_{\rm 2SDSS}$ and TiO$_1$ increase when [$\alpha$/Fe] increases, the TiO$_{\rm 2SDSS}$ decreases but TiO$_1$ increases for C18.  
These differences mean that, for given age, [M/H], [$\alpha$/Fe] and IMF, the MILES and C18 SSPs predict different line strengths.  As a result, when fitting to the line strengths we measure in MaNGA, MILES and C18 will return different SSP parameters.
Nevertheless, this exercise has shown that, at least for MILES, the subset of line combinations we use in the main text provide particularly simple routes to the SSP parameters of most interest.  

To explore alternatives, in Section~3.4 of the main text we considered the effect of also allowing C to be enhanced, by setting [C/Fe] equal to [$\alpha$/Fe].  We have also studied the effect of enhancing [Ti/Fe] over and above its contribution to [$\alpha$/Fe].  To illustrate, we assumed that the extra [Ti/Fe] enhancement equals [$\alpha$/Fe].  This is motivated by, e.g. Figure~13 in \cite{Conroy2014}, but is more extreme.  I.e., if [$\alpha$/Fe]=0.3, then this means that [O/Fe] = 0.3 as does [Mg/Fe] and similarly for all the other elements which appear in the top left panel of Figure~A2, except for [Ti/Fe], which equals 0.6 (the 0.3 of the other $\alpha$ elements plus the extra 0.3 enhancement because we assumed that the extra [Ti/Fe] enhancement equals [$\alpha$/Fe] and [$\alpha$/Fe] = 0.3).  As one might expect from Figure~A2, this has the effect of shifting the TiO2-[MgFe] grids shown in the main text upwards ([MgFe] is not sensitive to [Ti/Fe]), alleviating the need for very bottom heavy IMFs, but at the cost of shifting the H$_\beta$-[MgFe] grids upwards as well, so the ages are all increased.  There is essentially no change to [M/H], because Ti elements contribute less than one percent of the overall metallicity, so even enhancing Ti by a factor of 5 makes essentially no difference to [M/H]) or [$\alpha$/Fe].  Although we do not show it here, we have found that the net effect is to decrease M$_*/L$ values and gradients, but the ages are pushed to very large values.  This also means that age gradients are not allowed, making it more difficult to explain the strong [$\alpha$/Fe] gradient.

\subsection{Best-fit SSP parameters and predicted line strengths}
In the previous subsections, an SSP is characterized by its age, total metallicity [M/H] and $\alpha$-enhancement [$\alpha$/Fe], and these are estimated from the strengths of just four line-indices:  H$_\beta$, $\langle$Fe$\rangle$, TiO$_{\rm 2SDSS}$ and [MgFe].  As these line-indices are not the only options, we now discuss alternatives.

As the bottom panels of Figure~\ref{fig:responseIMF} show TiO$_1$ is sensitive to the IMF.  However, it is also sensitive to [$\alpha$/Fe], especially for MILES (c.f. Figure~\ref{fig:response}).
More importantly, TiO$_1$ is more sensitive to [C/Fe] than is TiO$_{\rm 2SDSS}$.  Therefore, if we wish to work with TiO$_1$, we may need to include [C/Fe] enhancement as an extra parameter, and account self-consistently for its effect on [M/H].  This is even more true for Mg4780 (which also depends strongly on [$\alpha$/Fe]).  Likewise, NaD responds strongly to [Na/Fe] and [$\alpha$/Fe], so to use it, we would need to include [Na/Fe] as a free parameter.  Since we enhance neither [C/Fe] nor [Na/Fe], we do not expect to find a good match to these indices.  Indeed, although we do not show it here, the measured TiO$_1$ line strengths lie off the model grids for standard or fiducial choices of [$\alpha$/Fe], for both MILES and C18 \cite[this is also consistent with Fig.E1 in][]{LaBarbera2019}.\footnote{\cite{LaBarbera2019} explore models which are enhanced in [Na/Fe], finding that they are necessary to describe NaD, and improve agreement with Mg4780.}  

In contrast, the indices we do use are {\em in}sensitive to both [C/Fe] and [Na/Fe], and only $\langle$Fe$\rangle$ responds to [$\alpha$/Fe].  This allows a determination of age, [M/H] and [$\alpha$/Fe] with fewer assumptions.  In principle, one could mix and match enhancements in other elements until one finds consistency.  However, this typically does not solve the problem that the C18 models tend to be too old (sometimes unacceptably old) and metal poor compared to MILES.

\begin{figure}
  \centering
  \includegraphics[width=0.9\linewidth]{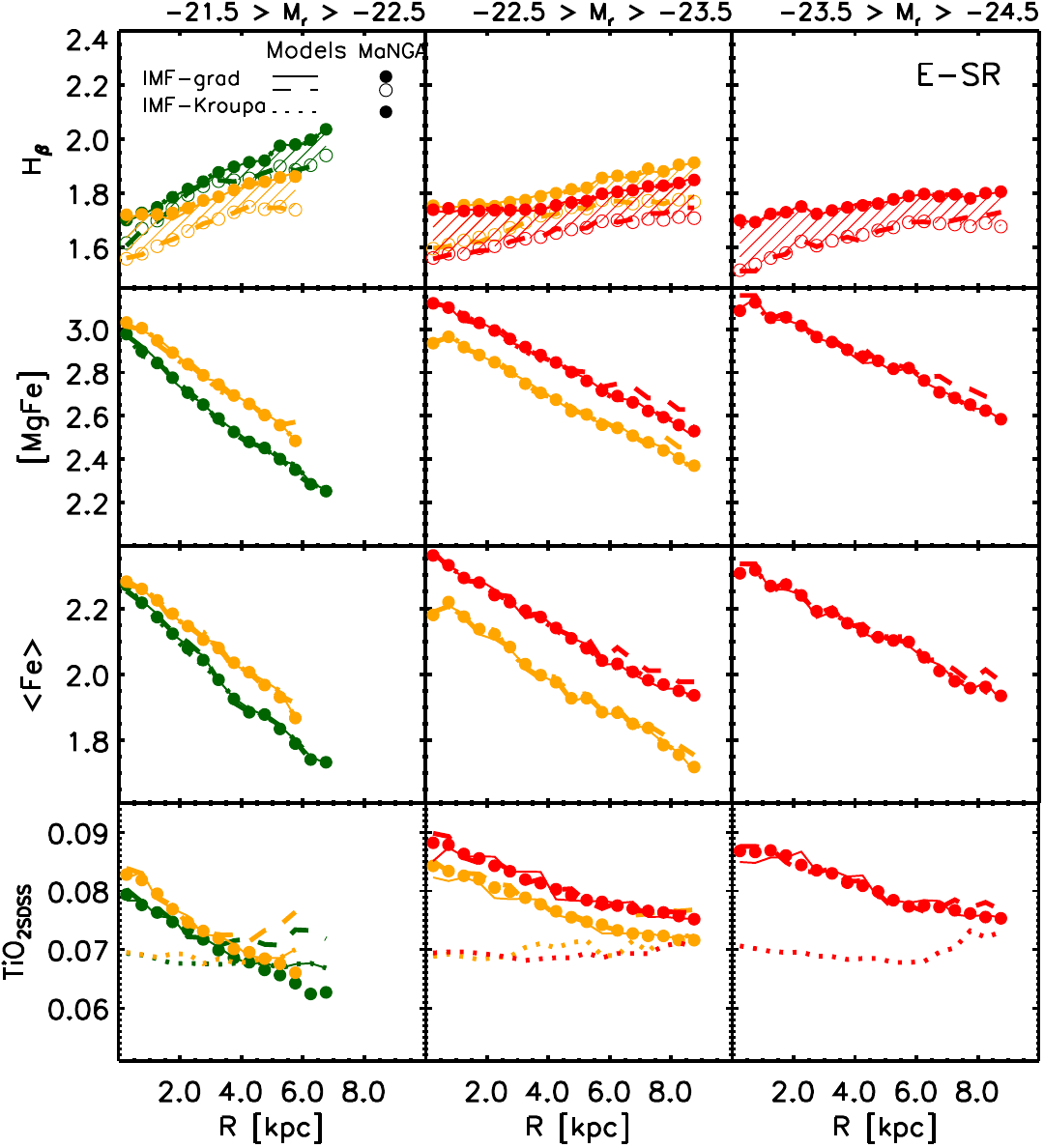}
  \caption{Comparison of measured line strengths for MaNGA E-SRs (filled symbols) with the values predicted by the MILES models when the SSP parameters are those shown in Figure~\ref{fig:miles}.  Open symbols in top panels show the measurements associated with the second of the two emission line corrections discussed in the main tex.  Models in which the IMF is fixed to Kroupa (dotted) or can vary (solid and dashed) describe the H$_\beta$, $\langle$Fe$\rangle$, and [MgFe] index strengths well, but SSPs with a Kroupa IMF are unable to produce the observed TiO$_{\rm 2SDSS}$ strengths (bottom panels).   
    Results for E-FRs and S0s are similar. }
  \label{fig:predict}
\end{figure}

We close with a check of the self-consistency of our approach.  Symbols in the different panels of Figure~\ref{fig:predict} show radial profiles of various measured line-strengths, and curves show the predicted profile if the IMF is fixed to Kroupa (dotted) or is allowed to vary (solid or dashed).  The top three rows show that the IMF choice doesn't matter much for H$_\beta$, [MgFe] and $\langle$Fe$\rangle$, but the bottom row shows that a Kroupa IMF cannot explain the TiO$_{\rm 2SDSS}$ line strength, whereas a varying IMF resolves the discrepancy.  Open circles in the top panels show the measurements, and dashed curves the associated (varying IMF) prediction, when the second of our two emission corrections is applied.

\section{Predicted broadband colors}\label{sec:bbcolors}

\begin{figure}
    \centering
    \includegraphics[width=0.8\linewidth]{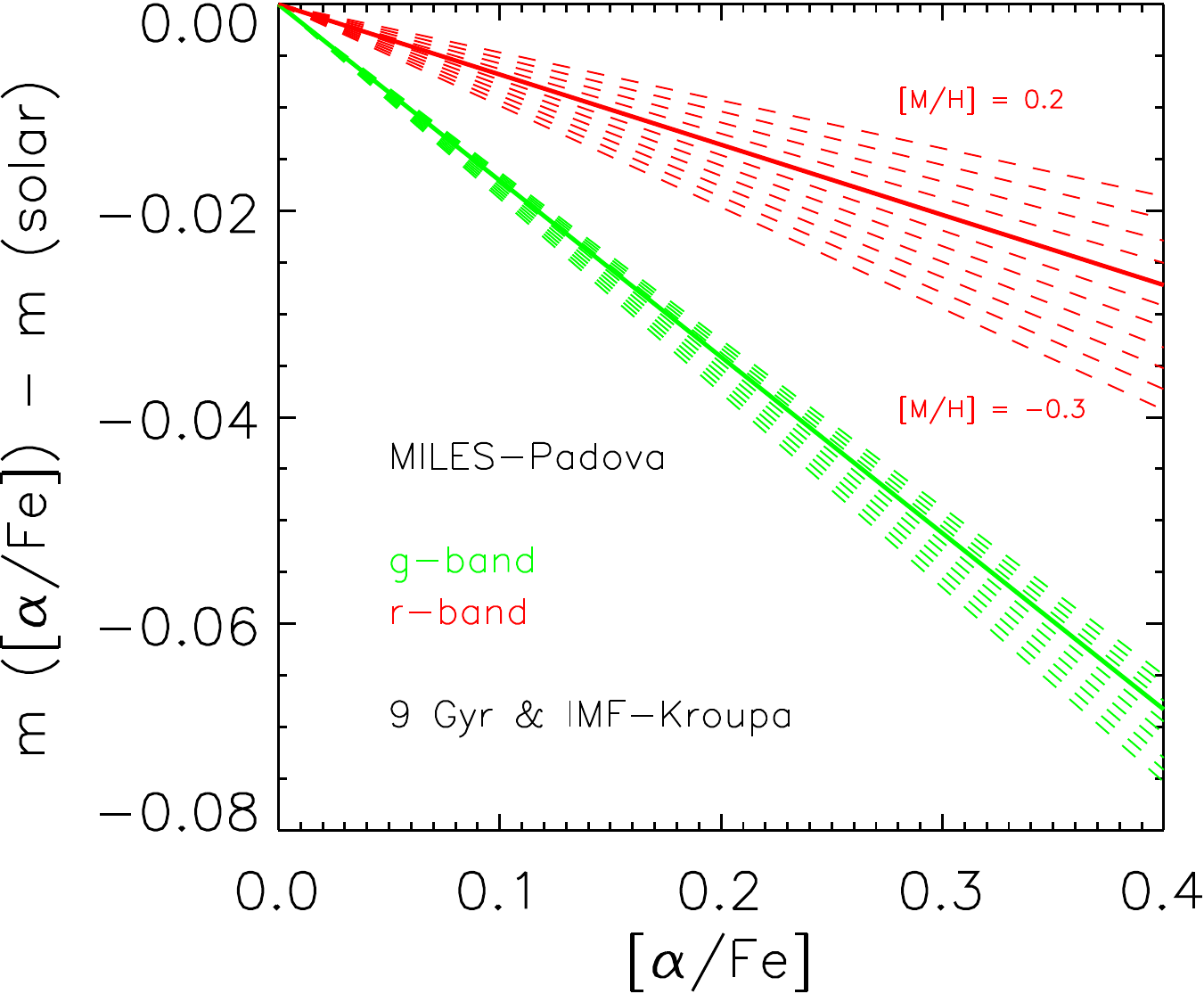}
    \includegraphics[width=0.8\linewidth]{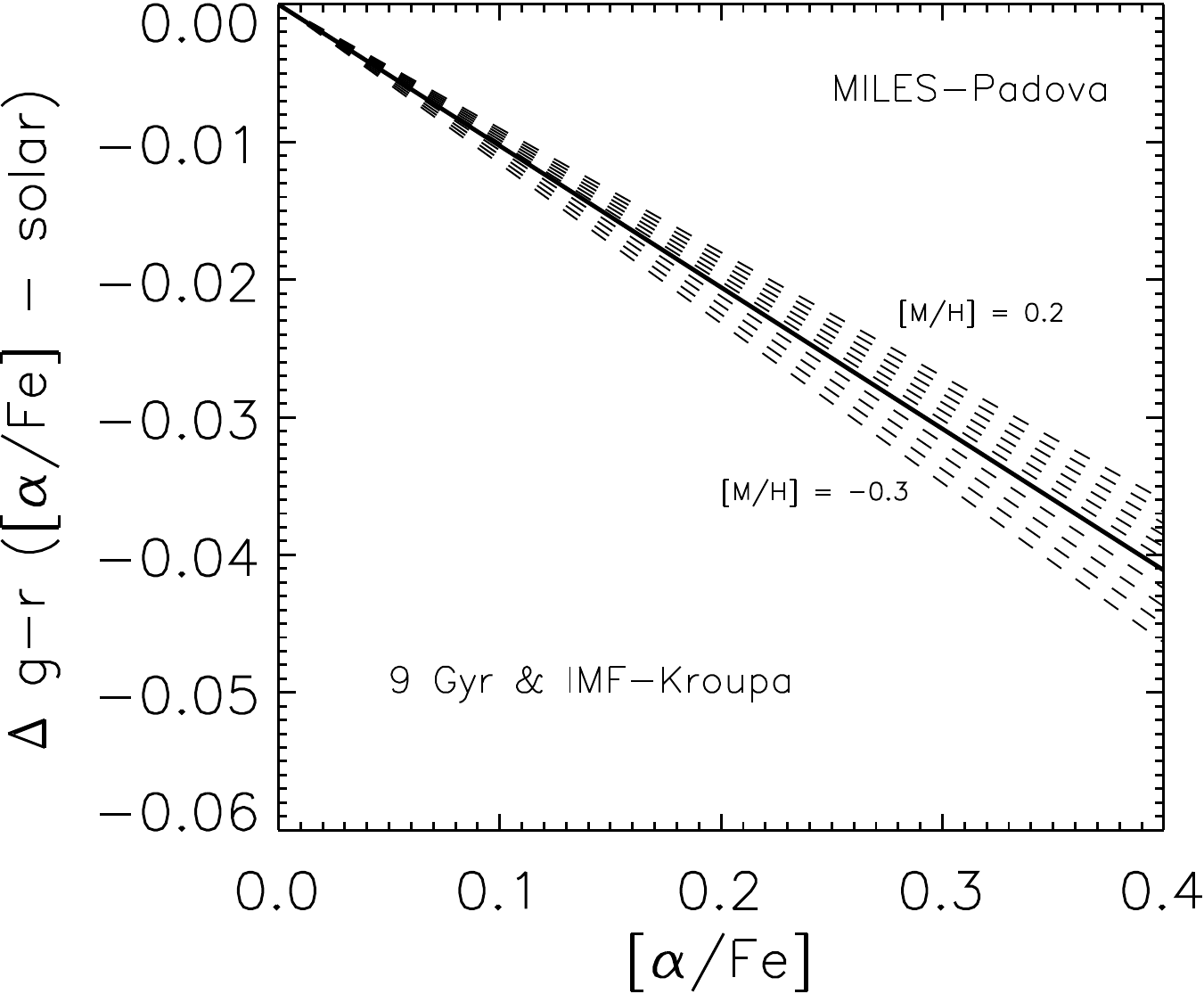}
    \caption{Dependence of $g$ and $r$ band luminosity on [$\alpha$/Fe] for a range of [M/H] when the age is 9~Gyrs and the IMF is Kroupa (top).  The difference is small, so it matters for colors (bottom), but not for $M_*/L$ ratios.}
    \label{fig:gr-alpha}
\end{figure}

In the main text, we used a few Lick indices to infer SSP parameters.  At face value, this ignores information in the rest of the spectrum.  If the SSP model is able to correctly predict the broad-band optical colors, then this is a nice consistency check.  

Performing this check is slightly involved because we wish to study how colors respond to both [$\alpha$/Fe] and IMF (in addition to age and metallicity), and this means we must be able to compute $L_g$ and $L_r$.  We do so by noting that the four SSP parameters define a spectrum:  To get $L_g$ or $L_r$, we simply integrate over this spectrum using the SDSS $g$- or $r$-band filter functions.  For the C18 models, we can enhance the spectra as we wish.  But for the MILES+Padova models, which are not $\alpha$-enhanced, we must scale the MILES+BaSTI models.  In practice, we scale the MILES+BaSTI $L_g$ and $L_r$ values similarly to how, following \cite{DS2019}, we scale the absorption line strengths.  In practice, because the MILES spectra are only provided in a few discrete bins in SSP parameter values, we interpolate to obtain the $L_g$ and $L_r$ values at the SSP values we want.  

The top panel of Figure~\ref{fig:gr-alpha} shows how the predicted brightness varies with [$\alpha$/Fe] for a range of [M/H] in the MILES+Padova models when the age is 9~Gyrs and the IMF is Kroupa. While there are obvious trends, they are smaller than 0.1~mags over the entire range of [$\alpha$/Fe] of interest, so they matter little for $M_*/L_r$ and $M_*/L_g$.  However, these trends are significant for the $g-r$ color; the bottom panel shows that [$\alpha$/Fe]$\sim 0.3$ can produce $g-r$ differences that are of order 0.03~mags different from when [$\alpha$/Fe]= 0. 

\begin{figure}
    \centering
    \includegraphics[width=0.8\linewidth]{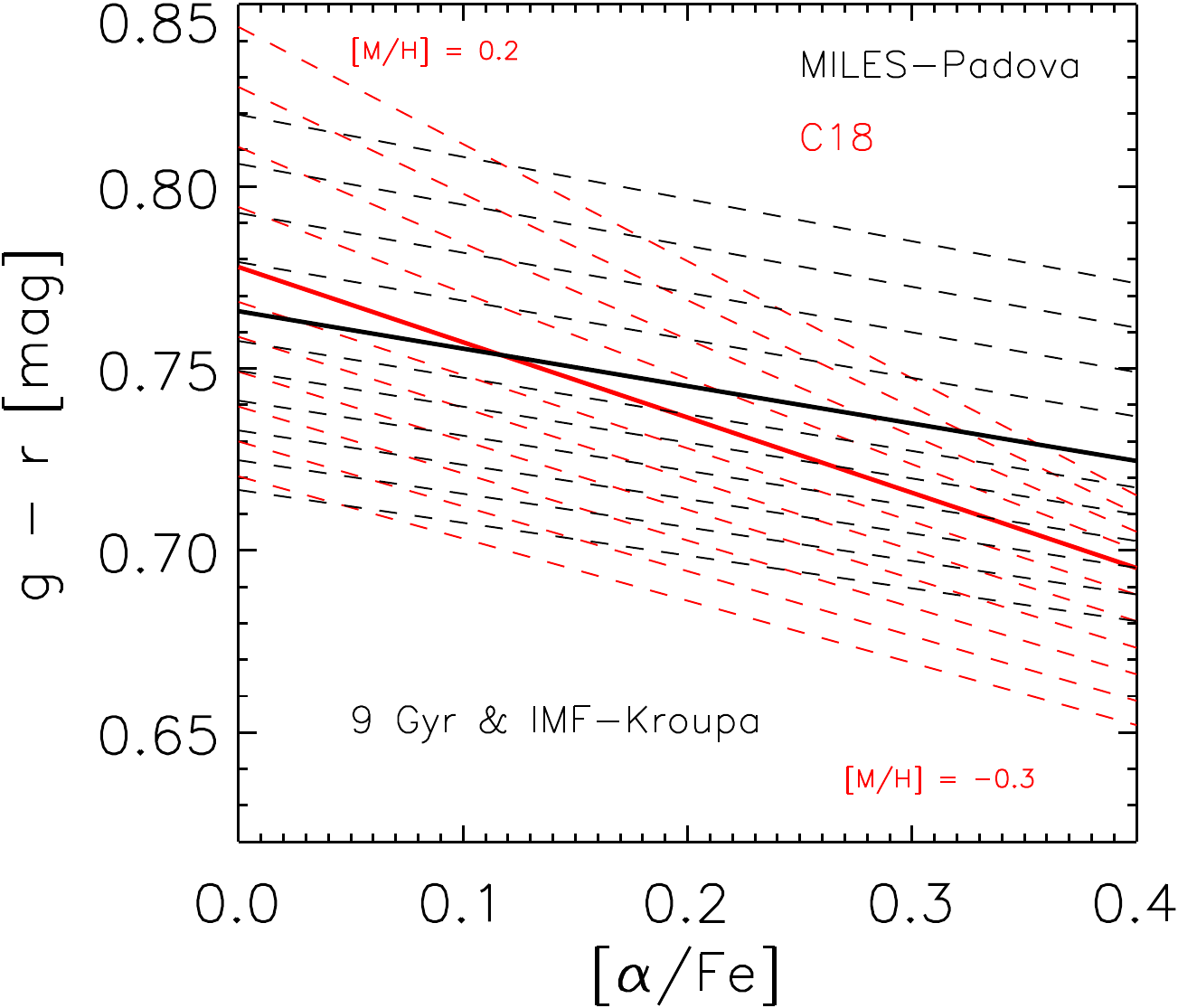}
    \caption{Dependence of $g-r$ color on [$\alpha$/Fe] for the MILES and C18 models for a 9~Gyr population, a Kroupa IMF, and a range of total metallicity values; bold lines are for [M/H]=0.  Increasing [M/H] makes objects redder, whereas increasing [$\alpha$/Fe] makes them slightly bluer, although this trend is stronger for C18.  Increasing [M/H] as one increases [$\alpha$/Fe] can leave the color unchanged.
    }
    \label{fig:gmrAlpha}
\end{figure}

Figure~\ref{fig:gmrAlpha} explores this further.  It shows how $g-r$ responds to changes in [$\alpha$/Fe] for the MILES and C18 models for a 9~Gyr population, a Kroupa IMF, and a range of [M/H] values.  Increasing [M/H] makes objects redder, whereas increasing [$\alpha$/Fe] makes them slightly bluer, although this trend is stronger for C18.  As a result, increasing [M/H] as one increases [$\alpha$/Fe] can leave the color unchanged.  (Note that we work with total metallicity [M/H]; we have checked that we reproduce the C18 trends in \citealt{Choi2019} if we were to fix [Fe/H]=0.)

Figure~\ref{fig:gmrAlpha} shows that the MILES and C18 models can predict rather different colors, especially at large [$\alpha$/Fe] and large [M/H].  However, most of the objects in the main text lie in the region where the two are in better agreement ([$\alpha$/Fe]$\sim 0.2$ and [M/H] that is slightly less than 0).  

\begin{figure}
    \centering
        \includegraphics[width=0.8\linewidth]{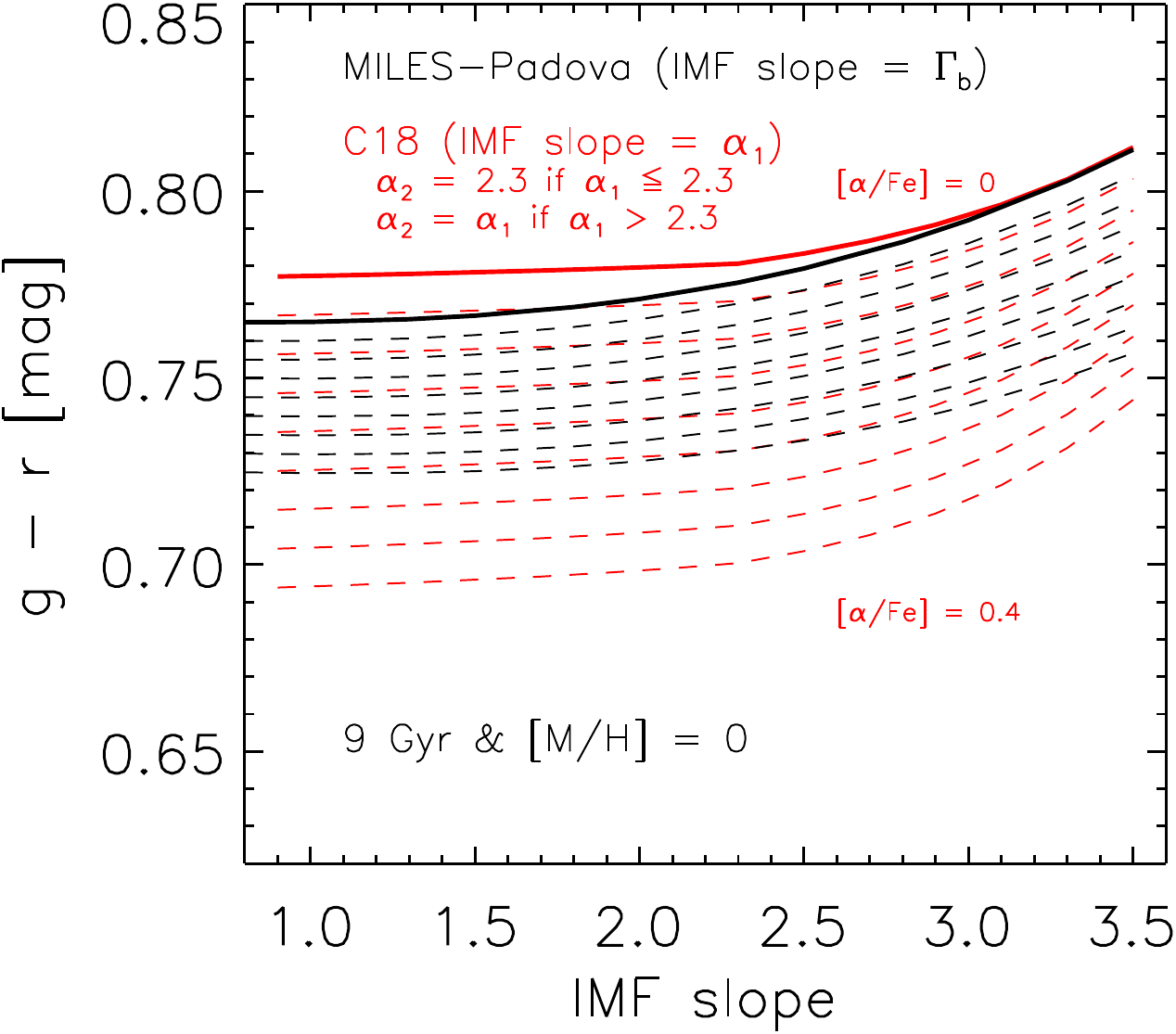}
    \caption{Dependence of $g-r$ color on IMF for the MILES and C18 models for a 9~Gyr population with [M/H]=0, and a range of [$\alpha$/Fe] values; bold lines are for [$\alpha$/Fe]=0.  For a given age, [M/H] and [$\alpha$/Fe], color is insensitive to IMF except when the IMF is extremely bottom heavy (large values of the `slope'), in which case larger slopes produce redder colors.  Increasing [$\alpha$/Fe] as the IMF slope increases can leave the color unchanged.
    }
    \label{fig:gmrIMF}
\end{figure}

Figure~\ref{fig:gmrIMF} shows a similar study of how $g-r$ responds to changes in the IMF, for the MILES and C18 models, for a 9~Gyr population with [M/H]=0, and a range of [$\alpha$/Fe] values.  Evidently, for a given age, [M/H] and [$\alpha$/Fe], color is insensitive to IMF (in ageement with previous work) except when the IMF is extremely bottom heavy (large values of the `slope'; see Figure~\ref{fig:imfs} for the correspondence between slope and IMF shape), in which case larger slopes produce redder colors.  Increasing [$\alpha$/Fe] as the IMF slope increases can leave the color unchanged.  Once again, we see that MILES and C18 predictions differ most at larger [$\alpha$/Fe] values than are typical of the MaNGA sample.

The weakness of these trends, and the obvious degeneracies are why $g-r$ is not a good diagnostic of either [$\alpha$/Fe] or the IMF. This is in addition to the fact that optical colors suffer from an age-[M/H] degeneracy, the breaking of which is why Lick indices were developed in the first place.

Having shown how $g-r$ responds to SSP parameters, Section~\ref{sec:colors} in the main text compares the colors predicted by our MILES+Padova SSPs with those observed.


\bsp	
\label{lastpage}
\end{document}